\documentclass[12pt,a4paper]{article}
\usepackage{amssymb,cite,graphics}
\newcommand\blank[1]{}

\newcommand{\resection}[1]{\setcounter{equation}{0}\section{#1}}
\newcommand{\appsection}[1]{\addtocounter{section}{1}
\setcounter{equation}{0} \section*{Appendix \Alph{section}~~#1}}

\begin{document}
\begin{titlepage}
\vskip 0.5cm
\begin{flushright}
DTP/00/33  \\
{\tt hep-th/0004121} \\
\end{flushright}
\vskip 1.5cm
\begin{center}
{\Large{\bf
 On the quantum reflection factor for the sinh-Gordon model with general
boundary conditions
}}
\end{center}
\vskip 0.8cm
\centerline{A. Chenaghlou
\footnote{{\tt alireza.chenaghlou@durham.ac.uk}}}
\vskip 0.9cm
\centerline{\sl\small Dept.~of Mathematical Sciences,
University of Durham, Durham DH1 3LE, UK\,}
\vskip 1.25cm
\begin{abstract}
\noindent
The one loop quantum corrections to the classical reflection factor
of the
sinh-Gordon model are calculated partially for general boundary
conditions. The model is
studied under boundary conditions which are compatible with 
integrability,
and in the framework of the conventional perturbation theory 
generalized
to the affine Toda field theory. It is found that the general form 
of the 
related quantum corrections are hypergeometric functions. 
\end{abstract}
\end{titlepage}
\setcounter{footnote}{0} 
\def\thefootnote{\fnsymbol{footnote}}

\resection{Introduction}

Affine Toda field theory \cite{AFZ,MOP} is an integrable quantum
field theory in
two-dimensional Minkowski space-time which possesses remarkable
properties and rich algebraic structure (for a review see \cite{C1}).
This theory becomes more interesting
\cite{CDRS,CDR,BCDR,FS,FK1,FK2} when it is restricted to a 
half-line. 
However, for most of the Toda
theories corresponding to affine simply-laced algebras, the boundary
conditions are  limited to a finite number which 
preserve
integrability. Corrigan et.al \cite{CDRS,BCDR,CDR} have classified
the boundary conditions which preserve classical integrability.
 However, there
still remains much to be studied in relation to quantum
integrability
on the half-line.  For  models based on $a_{n}^{(1)}$
much is now known 
\cite{CDRS,Ga,DG,PB}. 
 The simplest affine Toda theory,  the sinh-Gordon model has been
studied much more than other models in the context of integrable
boundaries. 
 This model is the only theory in the $ade$ series of affine 
Toda field theory for which continuous boundary parameters
are possible. 

In recent years there has been considerable  interest
\cite{PB,C,K1,K2,T,BCDS}
 in
perturbative affine Toda
field theory. The motivation behind this fact is that the
boundary
S-matrices of the models are largely unknown. The most 
progress
has been made for $a_{1}^{(1)}$ affine Toda field theory for which 
the general form of the boundary S-matrix has been found by
Ghoshal \cite{G}. 
In fact, the boundary bootstrap equations yield the boundary
 S-matrices up to some unknown parameters. The perturbation 
method not only provides an additional check  of the results
which come from the bootstrap technique, but also it could 
make a connection between the unknown parameters of the boundary
S-matrices and the boundary parameters which are involved
in the Lagrangian formulation of the theories.

Firstly, Ghoshal and Zamolodchikov \cite{GZ} obtained the
soliton reflection factors in the sine-Gordon model 
with a boundary consistent with integrability. 
Then, Ghoshal \cite{G} using these results calculated the
reflection factors of the soliton-anti-soliton bound states 
(the breathers) of the model. One of the interesting problem 
in the boundary sine(sinh)-Gordon model is to find the relation
between the free parameters appearing in  Ghoshal's formula 
and the boundary data appearing in the Lagrangian of the model. 
 Corrigan \cite{C} was the first to notice  that the lightest
breather reflection
factor of the sine-Gordon model is identical to the reflection
factor of the
sinh-Gordon model after an analytic continuation in the
coupling constant. In a recent paper Corrigan and Delius 
\cite{CD} studied the boundary breather states of the
sinh-Gordon model on a half-line. They calculated the energy 
spectrum of the boundary states in two ways, by using the 
bootstrap equations then by using a WKB approximation. 
By comparing the results obtained by the two methods,   
they provided  strong evidence for a conjectured relationship 
between the boundary parameters, the bulk coupling constant and 
the parameters appearing in the quantum reflection factor 
calculated by Ghoshal.  They carried out the
calculations in the special case  when the boundary
parameters are equal and the boundary condition preserve the 
$\phi \rightarrow -\phi$ symmetry of the bulk theory. 

In \cite{C} the quantum corrections  up to
$O(\beta^{2})$ to the classical
reflection factor of the sinh-Gordon model 
were found 
  when the boundary
parameters are equal. In this case, the static background 
configuration
is $\phi=0$. If the boundary data are different then, the 
lowest
energy solution will not be a trivial background. The
corresponding
perturbation theory involves complicated coupling constants and
two-point  Green function as well. Recently \cite{CC}
the quantum reflection
factor has been calculated in one loop order up to the
first
order in
the difference of the two boundary parameters. 
The result \cite{CC} provide a further verification of Ghoshal's
formula. This  paper extends the results of \cite{CC}, by
calculating the quantum reflection factor for any value of the
boundary parameters. It is found that most part
of the related quantum corrections to the classical reflection factor
may be expressed in terms of  hypergeometric functions. 
This result and how it could relate  to Ghoshal's formula is discussed in
the conclusions.
\resection{Boundary sinh-Gordon model}

The sinh-Gordon theory on the half-line is a massive scalar quantum
field theory
in 1+1 dimension whose corresponding untwisted affine Kac-Moody 
algebra is
$a_{1}^{(1)}$. The Lagrangian density of the theory is:
\begin{equation}
\bar{\mathcal{L}}=\theta(-x) \mathcal{L} -\delta(x) \mathcal{B}
\end{equation}
Here, $ \mathcal{L}$ is the bulk Lagrangian density of the model 
which is
given by 
\begin{eqnarray}
\mathcal{L}&=&\frac{1}{2}\partial_{\mu}\phi\partial^{\mu}\phi -
V(\phi)
\nonumber\\
&=&\frac{1}{2}\partial_{\mu}\phi\partial^{\mu}\phi
-\frac{2m^{2}}{\beta^{2}}\cosh(\beta \alpha \phi),
\end{eqnarray}
where m and $\beta$ are a mass scale and a coupling constant of the
theory. Moreover, the boundary potential $\mathcal{B}$ has the  
generic
form \cite{GZ}
\begin{equation}
\mathcal{B}=\frac{m}{\beta^{2}}\left(\sigma_{0}e^{-\frac{\beta}
{\sqrt{2}}\phi}+
\sigma_{1}e^{\frac{\beta}{\sqrt{2}}\phi} \right).
\end{equation}
In the above relation, the two real coefficients $\sigma_{0}$ and
$\sigma_{1}$ are arbitrary and indicate \cite{CDR,FS} the degrees  of
freedom allowed at
the boundary. In fact, Bowcock et.al \cite{BCDR}  obtained some
results about the form of the  the
boundary term via a generalized Lax pair when there is a boundary.
For
further discussion on the boundary parameters see \cite{CDR,FS}.

The sinh-Gordon model is integrable classically which means there are
infinitely many independent conserved quantities $Q_{\pm s}$ where 
s is an
arbitrary odd integer. On the other hand, the model is integrable 
after
quantizing which implies the S-matrix describing the n-particles
scattering factorises into a product of two-particles scattering
amplitudes. The S-matrix describing the elastic scattering of two
sinh-Gordon particles of relative rapidity $\theta$ is conjectured 
to have
the form \cite{FK,ZZ,AFZ} 
\begin{equation}
S(\theta)=-\frac{1}{(B)(2-B)}
\end{equation}  
where we use  the hyperbolic building  blocks 
\begin{equation}
(x)=\frac{\sinh(\theta/2+ \frac{i\pi
x}{4})}{\sinh(\theta/2-\frac{i\pi
x}{4})},
\end{equation}
 and the quantity B is related to the coupling constant $\beta$ by
$B= \frac{2\beta^{2}}{4\pi+\beta^{2}}$.

In order to maintain the integrability on the half-line, the boundary
potential must satisfy the following equation
\begin{equation}
\frac{\partial\phi}{\partial x}=-\frac{\partial\mathcal{B}}
{\partial\phi}
 \hspace{.25in}   \hbox{at}\,\,\,\,  x=0,  
\end{equation}  
or
\begin{equation}
\frac{\partial \phi}{\partial x}=-\frac{\sqrt{2}m}{\beta} \left(
\sigma_{1}
e^{\beta \phi/\sqrt{2}} - \sigma_{0} e^{- \beta \phi/\sqrt{2}} 
\right)
\hspace{.25in}      \hbox{at}      \,\,\,\,     x=0,
\end{equation}
where we  use the normalization condition $\alpha^{2}=2$ 
which is
customary in affine Toda field theory. In what follows the 
dimensional
mass parameter $m$ will be taken to unity. It is also convenient
to use $\sigma_{i}=\cos a_{i}\pi$. For the boundary
sinh-Gordon
model, besides to the two-particle S-matrix it is necessary to know 
the
boundary S-matrix or reflection factor describing one particle 
reflection
off the boundary. Firstly, Ghoshal and Zamolodchikov
\cite{GZ} calculated the
soliton reflection factors for the sine-Gordon model by  solving the
boundary Yang-Baxter equation. Then, Ghoshal
\cite{G} calculated the soliton-antisoliton bound state reflection
factor. He used
the boundary bootstrap equations along with the result of 
reference \cite{GZ}. The
general form of the quantum reflection factor in sinh-Gordon model 
may be
derived by regarding the lightest breather reflection factor of the
sine-Gordon model \cite{G}, calculated by Ghoshal, and performing
analytic continuation in the coupling
constant to find
\begin{equation}
K_{q}(\theta)=\frac{(1)(2-B/2)(1+B/2)}{(1-E(\sigma_{0},\sigma_{1},
\beta))
(1+E(\sigma_{0},\sigma_{1},\beta) )(1-F(\sigma_{0},\sigma_{1},\beta)
)(1+F(\sigma_{0},\sigma_{1},\beta) )}.
\end{equation}
Note the bulk reflection symmetry leads to $F=0$ or $E=0$ when
$\sigma_{0}=\sigma_{1}$ (note only one of them vanishes).
In fact, the exact form of the $E$ and $F$ is an open and hard 
problem. Recently
Corrigan and Delius \cite{CD} obtained the function E in the
special case when
$\sigma_{0}=\sigma_{1}=\cos a\pi$ and $\frac{1}{2}<a<1$ as
\begin{equation}
E=2a(1-B/2).
\end{equation}
They found the above formula by equating the results of the  WKB
approximation 
method and
the bootstrap technique. 

\resection{Low order perturbation theory}

 For  affine Toda field theory the perturbative 
calculation is
performed around the static background field configuration,  so 
standard Feynman Rules may be used. By  expanding  the bulk 
and
boundary potentials in terms of the coupling constant $\beta$,  
the three and four point couplings can be deduced. We find for the 
sinh-Gordon
theory
\begin{equation}
C_{bulk}^{(3)}=\frac{2\sqrt{2}}{3}
\beta\sinh(\sqrt{2}\beta \phi_{0})
\end{equation}

\begin{equation}
C_{bulk}^{(4)}= \frac{1}{3}
\beta^{2}\cosh
(\sqrt{2}\beta \phi_{0}),
\end{equation}
where $\phi_{0}$ represent the background solution to the equation of
motion of the model and similarly
\begin{equation}
C_{boundary}^{(3)}=\frac{\sqrt{2} 
\beta}{12}\left(
\sigma_{1}e^{\beta \phi_{0}/\sqrt{2}}-\sigma_{0}e^{-\beta
\phi_{0}/\sqrt{2}}\right)
\end{equation}
\begin{equation}
C_{boundary}^{(4)}=\frac{\beta^{2}}{48}\left(\sigma_{1}
e^{\beta
\phi_{0}/\sqrt{2}}+\sigma_{0} e^{-\beta\phi_{0}/\sqrt{2}}
\right).
\end{equation}
On the other hand, the static background field can be found through
linear perturbation of the equation of motion and the boundary
condition of the model \cite{CDR,C} to
obtain
\begin{equation}\label{background}
e^{\beta \phi_{0}/\sqrt{2}}=\frac{1+e^{2(x-x_{0})}}{1-e^{2(x-x_{0})}},
\end{equation}
where the parameter $x_{0}$ is related to the boundary parameters  by
\begin{equation}\label{boundary parameter}
\coth x_{0}=\sqrt{\frac{1+\sigma_{0}}{1+\sigma_{1}}}.
\end{equation}
So, the three and four point couplings corresponding to the bulk 
potential
take the forms
\begin{equation}
C_{bulk}^{(3)}=\frac{4\sqrt{2}}{3} \beta\cosh 2(x-x_{0})\left(
\coth^{2}2(x-x_{0}) -1\right),
\end{equation}

\begin{equation}
C_{bulk}^{(4)}=\frac{1}{3} \beta^{2}\left( 2 \coth^{2}
2(x-x_{0})-1 \right).
\end{equation}
In the same manner the three and four point couplings of the 
boundary are
given by
\begin{equation}
C_{boundary}^{(3)}=\frac{\sqrt{2} 
\beta}{12}\left(
\sigma_{1}\coth x_{0}-\sigma_{0}\tanh x_{0}\right),
\end{equation}
\begin{equation}
C_{boundary}^{(4)}=\frac{\beta^{2}}{48}\left(\sigma_{1}
\coth x_{0} +\sigma_{0}\tanh x_{0}\right).
\end{equation}

The next step is to find the propagator for the theory.
It has
been shown \cite{C} that the two-point Green function for the
sinh-Gordon model on
a half-line is 
\begin{eqnarray}
G(x,t;x',t')&=&\int \int \frac{d\omega}{2\pi} \frac{dk}{2\pi}
\frac{ie^{-i \omega (t-t')} }{\omega^{2}- k^{2}-4 +i \rho}
\left(f(k,x)f(-k,x') e^{ik(x-x')}
\right. \nonumber\\
& & \hspace{1.35in}+\left.K_{c}f(-k,x)f(-k,x')e^{-ik(x+x')}\right),
\end{eqnarray}
where
\begin{equation}
f(k,x)=\frac{ik-2 \coth 2(x-x_{0})}{ik+2}
\end{equation}
and $K_{c}$ is the classical reflection factor of the model which 
is equal
to 
\begin{equation}\label{classical reflection}
K_{c}=\left(\frac{(ik)^{2}+2ik\sqrt{1+\sigma_{0}}\sqrt{1+\sigma_{1}}
+2(\sigma_{0}+\sigma_{1})}{(ik)^{2}-2ik\sqrt{1+\sigma_{0}}
\sqrt{1+\sigma_{1}}+
2(\sigma_{0}+\sigma_{1})}\right)\left(\frac{ik-2}{ik+2}\right).   
\end{equation}
The  classical reflection factor (3.13) can be derived from the
quantum
reflection factor (2.8) by considering the classical limit i.e. when
$\beta \rightarrow 0$. Because, in this limit \cite{CDR} $E=
a_{0}+a_{1}$ and $F=a_{0}-a_{1}$. 
Now following the idea introduced by Kim \cite{K1} and developed by
Corrigan \cite{C}, we may
calculate the one loop quantum corrections to the classical reflection
factor after perturbation calculation of the two-point function and
then by finding the coefficient of $e^{-ik(x+x')}$ in the residue of
the on-shell pole in the asymptotic region $x,x'\rightarrow -\infty$.

In order to calculate the one loop  $\left(O(\beta^{2})\right)$
quantum corrections to the classical reflection factor, we use the
standard
perturbation theory which is generalized \cite{K1,K2,C,T} to the
affine Toda field theory on the half-line.
In general, at $\left(O(\beta^{2})\right)$
 there are three basic kinds of Feynman diagrams contribute
to the
two-point propagator of affine Toda field theory . These are shown in
figure 1.
Moreover, by inspection of the forms of the three point and four 
point
couplings which we have found, it is clear that all types of these
diagrams
are involved in our problem.
   
\begin{center}  
\includegraphics
{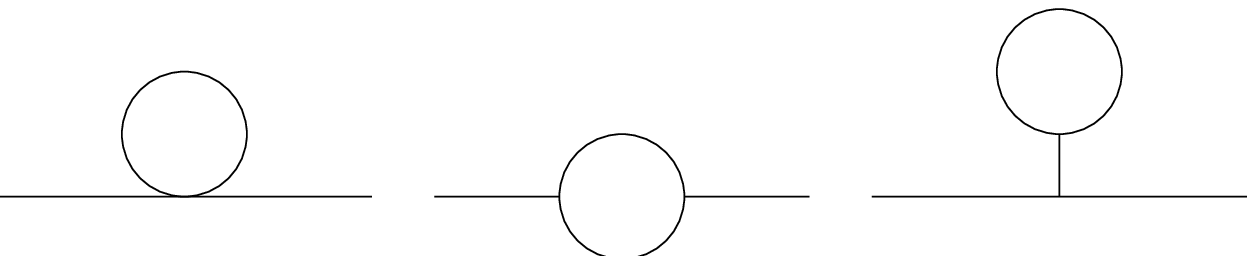}
\vspace{.25in}

\hspace{0in}I\hspace{1.6in} II\hspace{1.55in} III

\vspace{.25in}
\hspace{0in}Figuer 1: Three basic Feynman diagrams in one loop
order.
\end{center}
\vspace{.25in}

In fact, when the boundary parameters are not equal then, the 
calculations corresponding to the one loop order in the sinh-Gordon
model are lengthy and intricate. In the following sections we try to
compute the contributions of types I and III diagrams to the 
reflection
factor. The remaining diagrams will be treated elsewhere.  Meanwhile,
it is instructive to start with type III.

\resection{Type III diagram (boundary-boundary)}

In this section we shall  calculate the contribution of the 
type
III diagram to the reflection factor when both vertices are located 
at the
boundary $x=0$. 
We are led to the following integral 
\begin{eqnarray}\label{type three bb}
& &-\frac{\beta^{2}}{4}(\sigma_{1}\coth x_{0}-\sigma_{0}\tanh
x_{0})^{2}
\nonumber\\
& & \times \int \int dt
dt'G(x_{1},t_{1};0,t)G(0,t;0,t')G(0,t';0,t')G(0,t;x_{2},t_{2})
\end{eqnarray}
in which the combinatorial factor has been taken into account.

Let us start by looking at the loop propagator $G(0,t';0,t')$ 
which is 
equal to 
\begin{eqnarray}
G(0,t';0,t')&=&i \int \int \frac{d\omega'}{2\pi}\,\frac{dk'}{2\pi}
\,\frac{1}
{\omega'^{2}-k'^{2}-4+i\rho}
f(-k',0)
\nonumber\\
& &\hspace{1in}\times \biggl(f(k',0) +K'(k')f(-k',0)\biggr),
\end{eqnarray} 
where
\begin{equation}
f(k',0)=\frac{ik'+2\coth 2x_{0}}{ik'+2}
\end{equation}
and $K'(k')$ is the classical reflection factor (\ref{classical
reflection}). After some
manipulation, we obtain 

\begin{eqnarray}
G(0,t';0,t')&=&i \int \int \frac{d\omega'}{2\pi}\,\frac{dk'}{2\pi}
\,\frac{1}
{\omega'^{2}-k'^{2}-4+i\rho} 
\nonumber\\
& &\times\frac{2ik'\left(
ik'-2\coth 2x_{0}
\right)}{(ik')^{2}-2ik'\sqrt{1+\sigma_{0}}\sqrt{1+\sigma_{1}}+2(
\sigma_{0}+
\sigma_{1})}. 
\end{eqnarray}
 The above integral is clearly divergent however, the divergence
can be removed by the infinite renormalization of the boundary
term. 
In other words, considering the following relation
\begin{eqnarray}
& &\frac{2ik'\left(
ik'-2\coth 2x_{0}
\right)}{(ik')^{2}-2ik'\sqrt{1+\sigma_{0}}\sqrt{1+\sigma_{1}}+2(
\sigma_{0}+
\sigma_{1})}
\nonumber\\
&
&\hspace{.2in}=2+4\frac{ik'\left(\sqrt{1+\sigma_{0}}\sqrt{1+
\sigma_{1}}
-\coth 2x_{0}\right)
-(\sigma_{0}+\sigma_{1})}{(ik')^{2}-2ik'\sqrt{1+\sigma_{0}}\sqrt{1+
\sigma_{1}}
+2(\sigma_{0}+\sigma_{1})},
\end{eqnarray}
it is seen that a minimal subtraction of the divergent part 
can be made by adding an appropriate counter term to the boundary,
replace the logarithmically divergent integral by the finite part.  
Hence,

\begin{eqnarray}
G(0,t';0,t')&=&4i \int \int \frac{d\omega'}{2\pi}\,\frac{dk'}{2\pi}
\,\frac{1}
{\omega'^{2}-k'^{2}-4+i\rho}
\nonumber\\
&
&\times\frac{ik'\left(\sqrt{1+\sigma_{0}}\sqrt{1+\sigma_{1}}
-\coth 2x_{0}\right)
-(\sigma_{0}+\sigma_{1})}{(ik')^{2}-2ik'\sqrt{1+\sigma_{0}}\sqrt{1+
\sigma_{1}}
+2(\sigma_{0}+\sigma_{1})}.
\end{eqnarray}
The integration over $\omega'$ may be performed by closing the
contour
into the upper half-plane and collecting a pole at
$\omega'=\sqrt{k'^{2}+4}$  so that 
\begin{equation}\label{loop Green}
G(0,t';0,t')=2  \int \frac{dk'}{2\pi}\,\frac{1}
{\sqrt{k'^{2}+4}}
\,\frac{ik'\left(\sqrt{1+\sigma_{0}}\sqrt{1+\sigma_{1}}
-\coth 2x_{0}\right)
-(\sigma_{0}+\sigma_{1})}{(ik')^{2}-2ik'\sqrt{1+\sigma_{0}}\sqrt{1+
\sigma_{1}}
+2(\sigma_{0}+\sigma_{1})}.
\end{equation}
In order to integrate over  $k'$, as before, one chooses a  
contour in the upper half-plane, however due to the branch cut the 
contour
has to run around the cut line. Moreover we assume that the roots
of the 
denominator of the integrand i.e. 
$2\cos \frac{(a_{0} \pm a_{1})\pi}{2}$
are positive, otherwise we may close the contour in the lower 
half-plane.
Therefore,  (\ref{loop Green})   converts  to
\begin{equation}
  4\int_{2}^{\infty} \frac{dy}{2\pi}\,\frac{1}
{\sqrt{y^{2}-4}}
\,\frac{y\left(\sqrt{1+\sigma_{0}}\sqrt{1+\sigma_{1}}
-\coth 2x_{0}\right)
+(\sigma_{0}+\sigma_{1})}{y^{2}+2y\sqrt{1+\sigma_{0}}\sqrt{1+
\sigma_{1}}
+2(\sigma_{0}+\sigma_{1})}
\end{equation}      
and  the above integral gives  the following result
\begin{equation}
G(0,t';0,t')=-\frac{a_{0}}{2}\cot
a_{0}\pi-\frac{a_{1}}{2}\cot a_{1}\pi.
\end{equation}

Now it is convenient to calculate the time integral of the other middle
propagator in 
(\ref{type three bb})
which is equal to 
\begin{eqnarray}
\int dt'G(0,t;0,t') &=&\int\int\int dt'
\,\frac{d\omega}{2\pi}\,\frac{dk}{2\pi}\,e^{-i\omega(t-t')}\, 
\frac{i}
{\omega^{2}-k^{2}-4+i\rho} 
\nonumber\\
& &\hspace{.5in}\times\frac{2ik\left(
ik-2\coth 2x_{0}
\right)}{(ik)^{2}-2ik\sqrt{1+\sigma_{0}}\sqrt{1+\sigma_{1}}+2(
\sigma_{0}+
\sigma_{1})}.
\end{eqnarray}
Clearly, in the boundary-boundary contribution (\ref{type three
bb}), it is seen that
the $t'$
dependence is involved only in the propagator $G(0,t;0,t')$   
 so, the
integration over $t'$ gives us a Dirac delta function which means 
we must
substitute zero for  $\omega$  and hence
\begin{equation}
\int dt'G(0,t;0,t') =\int\frac{dk}{2\pi} \left(\frac{i}
{-k^{2}-4}\right) \left(\frac{2ik\left(
ik-2\coth 2x_{0}
\right)}{(ik)^{2}-2ik\sqrt{1+\sigma_{0}}\sqrt{1+\sigma_{1}}+2(
\sigma_{0}+
\sigma_{1})}\right).
\end{equation}
As we mentioned before, throughout this paper we assume that the 
roots
of
$P(k)=(ik)^{2}-2ik\sqrt{1+\sigma_{0}}\sqrt{1+\sigma_{1}}+2(
\sigma_{0}+\sigma_{1})$
which are equal to $2\cos\frac{(a_{0} \pm a_{1})\pi}{2}$ are 
positive, so the     
$P(k)$ has no pole in the upper half-plane. Obviously, if the roots 
are  negative
then we can choose the contour in the lower half-plane in which no 
pole is
inserted. Therefore, 
\begin{equation}
\int dt'G(0,t;0,t') = \frac{i(1+
\coth 2x_{0})}
{2+2\sqrt{1+\sigma_{0}}\sqrt{1+\sigma_{1}}+(\sigma_{0}+
\sigma_{1})}
\end{equation} 
and by substituting $\sigma_{0}=\cos a_{0}\pi$ and 
$\sigma_{1}=\cos a_{1}\pi$, we
obtain
\begin{equation}\label{Green 00}
\int dt'G(0,t;0,t') = -\frac{i}{4\cos\frac{a_{0}\pi}{2}
\cos\frac{a_{1}\pi}{2}}.
\end{equation}

Up to now, the boundary-boundary contribution has the  form
\begin{eqnarray}\label{boundary-boundary}
& &-\frac{i\beta^{2}}{32}\,\frac{(\sigma_{1}\coth
x_{0}-\sigma_{0}\tanh
x_{0})^{2}(a_{0}\cot a_{0}\pi+a_{1}\cot a_{1}\pi)}
{\cos \frac{a_{0}\pi}{2}\cos\frac{a_{1}\pi}{2}}
\nonumber\\
& &\times\int dt\int \int 
\frac{d\omega_{1}}{2\pi}\,\frac{dk_{1}}{2\pi}\,\frac{i
e^{-i\omega_{1}(t_{1}-t)}}{\omega_{1}^{2}-k_{1}^{2}-4+i\rho}
\biggl(
f(k_{1},x_{1})f(-k_{1},0)e^{ik_{1}x_{1}}
\biggr.\nonumber\\
& &\hspace{2in}+\biggl.K_{1}(k_{1})f(-k_{1},x_{1})f(-k_{1},0)
e^{-ik_{1}x_{1}}\biggr)
\nonumber\\
& &\times\int \int
\frac{d\omega_{2}}{2\pi}\,\frac{dk_{2}}{2\pi}\,\frac{i 
e^{-i\omega_{2}(t-t_{2})}}{\omega_{2}^{2}-k_{2}^{2}-4+i\rho}
\biggl(f(k_{2},x_{2})f(-k_{2},0)e^{-ik_{2}x_{2}}
\biggr.\nonumber\\
& &\hspace{2in}\biggl.+K_{2}(k_{2})f(-k_{2},x_{2})f(-k_{2},0)
e^{-ik_{2}x_{2}}\biggr).
\end{eqnarray}
First of all, it is necessary to perform the transformation 
$k_{1}\rightarrow
-k_{1}$ in the first term of the first propagator. Secondly, 
integration over $t$ ensures energy conservation at the interaction
vertex and 
generates a Dirac delta function because of which we can set 
$\omega_{1}=\omega_{2}$.
Moreover, it is better to define a new function as
\begin{equation}
A(k,x)=f(-k,x)f(k,0)+K(k)f(-k,x)f(-k,0)
\end{equation}
or, in an expanded form,
\begin{eqnarray}
A(k,x)&=&\frac{ik+2\coth 2x_{0}}{ik+2}
\,\,\frac{ik+2\coth 2(x-x_{0})}{ik-2}
\nonumber\\
& &+\frac{(ik+2\cos\frac{(a_{0}+a_{1})\pi}{2})(ik+2\cos\frac{(a_{0}-
a_{1})\pi}{2})}
{(ik-2\cos\frac{(a_{0}+a_{1})\pi}{2})(ik-2\cos\frac{(a_{0}-a_{1})\pi
}{2})}
\nonumber\\
& &\times\frac{ik-2\coth 2x_{0}}{ik-2}
\,\,\frac{ik+2\coth 2(x-x_{0})}{ik+2},
\end{eqnarray}
then,  the expression (\ref{boundary-boundary})  reduces to 
\begin{eqnarray}
& &-\frac{i\beta^{2}}{32}\,\frac{(\sigma_{1}\coth
x_{0}-\sigma_{0}\tanh
x_{0})^{2}(a_{0}\cot a_{0}\pi+a_{1}\cot a_{1}\pi)}
{\cos \frac{a_{0}\pi}{2}\cos\frac{a_{1}\pi}{2}}\nonumber\\
& &\times \int \int\int
 \frac{d\omega_{1}}{2\pi}\,\frac{dk_{1}}{2\pi}
 \,\frac{dk_{2}}{2\pi}
 \,e^{-i\omega_{1}(t_{1}-t_{2})}\,\frac{i}
{\omega_{1}^{2}-k_{1}^{2}-4+i
\rho}
 \,\frac{i}{\omega_{1}^{2}-k_{2}^{2}-4+i\rho}               
 \,e^{-ik_{1}x_{1}}
 \,e^{-ik_{2}x_{2}} \nonumber\\
& & \nonumber\\
& &\hspace{1in} \times A(k_{1},x_{1})\,A(k_{2},x_{2}).
\end{eqnarray}
Obviously, what we need to do next is to integrate over the momenta
$k_{1}$ and
$k_{2}$ and this   may be achieved by closing the contours  in the
upper
half-plane and considering the  poles at
$\hat{k}_{1}=k_{1}=k_{2}=\sqrt{\omega_{1}^{2}-4}$. Note, the 
additional
poles due
to functions $A(k_{1},x_{1})$ and $A(k_{2},x_{2})$ are not 
important because
their contributions will be exponentially damped as $x_{1},x_{2}
\rightarrow
-\infty$. Therefore, the boundary-boundary contribution is
\begin{eqnarray}
& &-\frac{i\beta^{2}}{32}\,\frac{(\sigma_{1}\coth
x_{0}-\sigma_{0}\tanh
x_{0})^{2}(a_{0}\cot a_{0}\pi+a_{1}\cot a_{1}\pi)}
{\cos \frac{a_{0}\pi}{2}\cos\frac{a_{1}\pi}{2}}
\nonumber\\
& &\times\int 
 \frac{d\omega_{1}}{2\pi}
 \,e^{-i\omega_{1}(t_{1}-t_{2})}\,e^{-i\hat{k}_{1}(x_{1}+x_{2})}
 \,\frac{1}{(2\hat{k}_{1})^{2}}
\,A(\hat{k}_{1},x_{1})\,A(\hat{k}_{2},x_{2}).
\end{eqnarray}

Now recall  the definition of the quantum reflection factor 
as  the
coefficient of $e^{-ik(x+x')}$ in the two-point Green function in 
the residue of
the on-shell pole when $x,x'\rightarrow
-\infty$. Thus, the
correction to the reflection factor from the type III (boundary-
boundary) piece
is
\begin{eqnarray}
& &-\frac{i\beta^{2}}{32}\,\frac{(\sigma_{1}\coth
x_{0}-\sigma_{0}\tanh
x_{0})^{2}(a_{0}\cot a_{0}\pi+a_{1}\cot a_{1}\pi)}
{\cos \frac{a_{0}\pi}{2}\cos\frac{a_{1}\pi}{2}}
\nonumber\\
& &\times\frac{1}{2\hat{k}_{1}} \left( \frac{(i\hat{k}_{1}+2\coth
2x_{0})^{2}}{(i\hat{k}_{1}+2)^{2}}
+2K(\hat{k}_{1})\frac{(i\hat{k}_{1}+2\coth 
2x_{0})(i\hat{k}_{1}-2\coth 2x_{0})}
{(i\hat{k}_{1}+2)^{2}}
\right.\nonumber\\
& &\hspace{2.5in}\left.+K^{2}(\hat{k}_{1})\frac{(i\hat{k}_{1}-2\coth
2x_{0})^{2}} {(i\hat{k}_{1}+2)^{2}} \right).
\end{eqnarray}

\resection{Type III (boundary-bulk)}

This section deals with the determination of the contribution of 
the type
III Feynman diagram to the classical reflection factor when one of 
the
vertices  corresponding to the loop is situated at the boundary and
the other one is inside the bulk
region. It is evident that in this case we have to
take into account the bulk three point coupling $C_{bulk}^{(3)}$ in 
the
corresponding vertex  as well as  the boundary three point 
coupling
$C_{boundary}^{(3)}$ in the other vertex. Meanwhile, the
combinatorial
factor associated with the related Feynman diagram must be 
considered as a
coefficient factor. Therefore, the contribution of the type III
(boundary-bulk) to the reflection factor may be written  as 
\begin{eqnarray}\label{boundary-bulk}
& &-2\beta^{2}(\sigma_{1}\coth x_{0}-\sigma_{0}\tanh x_{0})\int 
\int \int 
dt dt' dx G(x_{1},t_{1};x,t)G(x,t;0,t')
\nonumber\\
&
&\hspace{2.2in} \times 
G(0,t';0,t')G(x,t;x_{2},t_{2})\sinh(\sqrt{2}\beta
\phi_{0}).
\end{eqnarray}
The propagator  $G(0,t';0,t')$   corresponding to the loop has
been 
found in the 
previous  section and  is given by
\begin{equation}
G(0,t';0,t')=-\frac{a_{0}}{2}\cot a_{0}\pi
-\frac{a_{1}}{2}\cot a_{1}\pi.
\end{equation}

The calculation of the other middle propagator i.e. $G(x,t;0,t')$ 
is the
next  step and clearly, the $t'$ dependence in 
(\ref{boundary-bulk})  is
included only in this propagator. Hence, it is convenient to
compute the
following relation
\begin{eqnarray}
\int dt'G(x,t;0,t')&=&\int dt'\int \int
\frac{d\omega}{2\pi}\,\frac{dk}{2\pi}\,e^{-i\omega(t-t')}
\,\frac{i}{\omega^{2}-k^{2}-4+i\rho}\biggl(
f(k,x)f(-k,0)e^{ikx}
\biggr.\nonumber\\
& &\hspace{2in}\biggl. +K(k)f(-k,x)f(-k,0)e^{-ikx}\biggr).
\end{eqnarray} 
Integrating over  $t'$ generates a Dirac delta function and so,
\begin{eqnarray}
\int dt'G(x,t;0,t')
&=&i \int \frac{dk}{2\pi}\,
\frac{1}{(-k^{2}-4+i\rho)}\biggl(
f(k,x)f(-k,0)e^{ikx}
\biggr.\nonumber\\
& &\hspace{1.3in}\biggl.+K(k)f(-k,x)f(-k,0)e^{-ikx}\biggr)
\end{eqnarray} 
and the residue theorem gives 
\begin{equation}\label{Green x0}
\int dt'G(x,t;0,t')=\frac{ie^{2x}}{8} \left(c_{0}+c_{1}\coth 2(x-
x_{0})
 \right),
\end{equation}
where
\begin{equation}
c_{0}=-c_{1}=-\left(1+\tan^{2}\frac{(a_{0}+a_{1})\pi}{4}\right)
\left(1+\tan^{2}\frac{(a_{0}-a_{1})\pi}{4}\right).
\end{equation}
In order to check the above result, if we set $x=0$ in 
(\ref{Green x0})
 then it will be equal to  (\ref{Green 00}).

Up to now the type III (boundary-bulk) contribution take the 
following
form, of course, after integrating over $t$:
\begin{eqnarray}\label{type three boundary-bulk}
& &\beta^{2}c_{0}(\sigma_{1}\coth x_{0}-\sigma_{0} \tanh x_{0})
(a_{0}\cot a_{0}\pi+a_{1}\cot a_{1}\pi)
\nonumber\\
& &\times \int_{-\infty}^{0} dx\int \int
\frac{d\omega_{1}}{2\pi}\,\frac{dk_{1}}{2\pi}\,e^{-i
\omega_{1}(t_{1}-t_{2})}
\,\frac{i}{\omega_{1}^{2}-k_{1}^{2}-4+i\rho} \biggl(f(k_{1},x_{1})
f(-k_{1},x)
e^{ik_{1}(x_{1}-x)}
\biggr.\nonumber\\
&
&\hspace{2.5in}\biggl.+K_{1}(k_{1})f(-k_{1},x_{1})f(-k_{1},x)
e^{-ik_{1}(x+x_{1})}
\biggr) 
\nonumber\\
& &\times\left\{\frac{ie^{2x}}{8} \left(1-\coth 2(x-x_{0})
\right)
\sinh(\sqrt{2} \beta \phi_{0})\right\}
\nonumber\\
& &\times \int \frac{dk_{2}}{2\pi}
\,\frac{i}{\omega_{1}^{2}-k_{2}^{2}-4+i\rho}\biggl( f(k_{2},x)
f(-k_{2},x_{2})
e^{ik_{2}(x-x_{2})}
\biggr.\nonumber\\
&
&\hspace{2.3in}\biggl.+K_{2}(k_{2})f(-k_{2},x)f(-k_{2},x_{2})
e^{-ik_{2}(x+x_{2})}
\biggr).
\end{eqnarray}
By multiplying the two propagators in 
(\ref{type three boundary-bulk}) by 
each other, it is clear
that one obtains
four pole 
 pieces  
and, as far as the integration over $x$ is concerned, if we do the
integration over $x$ on one of them
 then, 
obviously the other three pole pieces could be done in the same 
manner. 
Hence, in what follows it 
is
sufficient  to treat only one of them and, meanwhile,
keeping 
those terms which are functions
 of $x$, we
are led to the following  complicated integral
\begin{equation}
\int_{-\infty}^{0}dx\,
\exp{\left\{2+i(k_{2}-k_{1})x\right\}} 
f(-k_{1},x)f(k_{2},x)
\sinh(\sqrt{2}\beta \phi_{0})
\biggl( 1-\coth
2(x-x_{0})
\biggr).
\end{equation}
After some  substitutions and collecting together powers
 of $\coth 2(x-x
_{0})$ we
obtain
\begin{eqnarray}\label{x integral}
& &\frac{1}{(ik_{1}-2)(ik_{2}+2)}\int_{-\infty}^{0}dx
\, \exp{\left\{2+i(k_{2}-k_{1})x\right\}}
\sinh(\sqrt{2}
\beta \phi_{0})
\biggl( -k_{1}k_{2}
\biggr.\nonumber\\
& &\hspace{.4in}\biggl.+(2ik_{2}-2ik_{1}+k_{1}k_{2})\coth 2(x-x_{0})
 +(2ik_{1}-2ik_{2}-4)\coth^{2} 2(x-x_{0})
\biggr.\nonumber\\
& &\hspace{3.7in}\biggl.+4\coth^{3} 2(x-x_{0}) \biggr).
\end{eqnarray} 
It is clear that in order to solve the above integral, it is
necessary to 
manipulate the following
integrals
\begin{equation}\label{Appendix A}
\int_{-\infty}^{0}dx\,
\exp{\left\{2+i(k_{2}-k_{1})x\right\}}\sinh(\sqrt{2}\beta 
\phi_{0})\coth^{n} 2(x-x_{0}), 
\end{equation}
where, $n=0,1,2,3$.

In fact in Appendix A, we have found the integrals (\ref{Appendix A}) 
  and
the
solutions of them are 
expressed in terms of hypergeometric functions. So, using the
 formulae in Appendix A and simplifying, we find that $\int$ in
(5.9) can be rewritten 

\begin{eqnarray}\label{F}
\mathcal{F}(k_{1},k_{2})&=&-\frac{3k_{1}k_{2}+4ik_{2}-4ik_{1}+8}{3}
\,\frac{1}{\sinh
2x_{0}}      
-\frac{3k_{1}k_{2}+6ik_{2}-6ik_{1}+13}{6}\,\frac{\cosh 2x_{0}}
{\sinh^{2}
2x_{0}}
\nonumber\\
& &-\frac{ik_{2}-ik_{1}+2}{3}\,\frac{\cosh^{2} 2x_{0}+1}{\sinh^{3}
2x_{0}}-\frac{1}{6}
\,\frac{\cosh^{3} 2x_{0}+5\cosh 2x_{0}}{\sinh^{4} 2x_{0}}
\nonumber\\
& 
&-\frac{12ik_{1}k_{2}-16k_{2}+16k_{1}+40i-(k_{2}-k_{1})(9k_{1}k_{2}+
16ik_{2}-16ik_{1}+34)}
{3(k_{2}-k_{1}-4i)}
\nonumber\\
& &\hspace{.5in}\times\,\, e^{-2x_{0}}F
\left(1,\frac{i}{4}(k_{2}-k_{1})+1,\frac{i}{4}(k_{2}-k_{1})+2,
e^{-4x_{0}} \right) 
\nonumber\\
& 
&-\frac{12ik_{1}k_{2}-48k_{2}+48k_{1}+136i-(k_{2}-k_{1})
(6k_{1}k_{2}+28ik_{2}-28ik_{1}+84)}
{3(k_{2}-k_{1}-8i)}
\nonumber\\
& &\hspace{.5in} \times \,\, e^{-6x_{0}}F
\left(2,\frac{i}{4}(k_{2}-k_{1})+2,\frac{i}{4}(k_{2}-k_{1})+3,
e^{-4x_{0}} \right)
\nonumber\\
& &+\frac{32k_{2}-32k_{1}-192i+(k_{2}-k_{1})(16ik_{2}-16ik_{1}+104)}
{3(k_{2}-k_{1}-12i)}
\nonumber\\
& &\hspace{.5in} \times \,\, e^{-10x_{0}}F
\left(3,\frac{i}{4}(k_{2}-k_{1})+3,\frac{i}{4}(k_{2}-k_{1})+4,
e^{-4x_{0}} \right)
\nonumber\\
& &+\frac{16k_{2}-16k_{1}-32i}
{(k_{2}-k_{1}-16i)}
\nonumber\\
& & \hspace{.5in} \times \,\, e^{-14x_{0}}F
\left(4,\frac{i}{4}(k_{2}-k_{1})+4,\frac{i}{4}(k_{2}-k_{1})+5,
e^{-4x_{0}} \right).
\end{eqnarray}
Now regarding  (\ref{type three boundary-bulk}), after 
doing
the transformation
$k_{1}\rightarrow
-k_{1}$
in the first term of the first propagator, all that remains is to 
integrate over the momenta
$k_{1}$ and $k_{2}$ and this can be achieved by closing the 
contours in the upper half-plane
and considering poles at $\hat{k}_{1}=k_{1}=k_{2}=
\sqrt{\omega_{1}^{2}-4}$.  The extra
poles in  the four functions $\mathcal{F}(\pm k_{1},\pm k_{2})$   are
not
important
because their contributions will be discounted when $x_{1}$ and 
$x_{2}$ go to $-\infty$.

Let us write down the type III (boundary-bulk) contribution to the 
reflection factor  
\begin{eqnarray}
& &\frac{\beta^{2}}{2}
\,\frac{\tan\frac{(a_{0}+a_{1})\pi}{4}\tan\frac{(a_{0}-a_{1})\pi}{4}}
{\cos\frac{a_{0}\pi}{2}\cos\frac{a_{1}\pi}{2}}
\,(a_{0}\cot a_{0}\pi+a_{1}\cot a_{1}\pi)
\nonumber\\
& &\times \int\frac{d\omega_{1}}{2\pi}\,e^{-i\omega_{1}(t_{1}-t_{2})}
\,e^{-i\hat{k}_{1}(x_{1}+x_{2})}
\,\frac{1}{(2\hat{k}_{1})^{2}}\,\frac{i\hat{k}_{1}+2\coth
2(x_{1}-x_{0})}{i\hat{k}_{1}-2}\,\frac{i\hat{k}_{1}+2\coth 2(x_{2}-
x_{0})}{i\hat{k}_{1}-2}
\nonumber\\
&
&\times 
\left\{\frac{i}{(i\hat{k}_{1}+2)^{2}}\mathcal{F}(-\hat{k}_{1},\hat{k}_{1})
-\frac{i}{(i\hat{k}_{1}+2)(i\hat{k}_{1}-2)}K_{1}(\hat{k}_{1})
\mathcal{F}(\hat{k}_{1},\hat{k}_{1})
\right.  \nonumber\\
&
&\hspace{.3in}\left.-\frac{i}{(i\hat{k}_{1}+2)(i\hat{k}_{1}-2)}
K_{1}(\hat{k}_{1})\mathcal{F}(-\hat{k}_{1},-\hat{k}_{1})
+\frac{i}{(i\hat{k}_{1}-2)^{2}}K_{1}^{2}(\hat{k}_{1})\mathcal{F}(\hat{k}_{1},
-\hat{k}_{1})\right\}
\end{eqnarray}
Now looking at the  function  $\mathcal{F}(k_{1},k_{2})$ given by
(\ref{F}), let us
show 
the
detailed forms  of 
 $\mathcal{F}(-\hat{k}_{1},\hat{k}_{1})$,
$\mathcal{F}(\hat{k}_{1},\hat{k}_{1})$,$\mathcal{F}(-\hat{k}_{1},-\hat{k}_{1})$
and $\mathcal{F}(\hat{k}_{1},-\hat{k}_{1})$. In fact, 
\begin{eqnarray}
\mathcal{F}(\hat{k}_{1},\hat{k}_{1})&=&-\frac{3\hat{k}_{1}^{2}+8}{3}
\,\frac{1}{\sinh
2x_{0}}+\frac{3\hat{k}_{1}^{2}+13}{6}\,\frac{\cosh 2x_{0}}
{\sinh ^{2}2x_{0}}
\nonumber\\
& &-\frac{2}{3}\,\frac{\cosh ^{2}2x_{0}+1}{\sinh
^{3}2x_{0}}-\frac{1}{6}\,\frac{\cosh
^{3}2x_{0}+5\cosh 2x_{0}}{\sinh ^{4}2x_{0}}
\nonumber\\
& &+\frac{3\hat{k}_{1}^{2}+10}{3}\,e^{-2x_{0}}\,F(1,1,2,e^{-4x_{0}})
\nonumber\\
& &+\frac{3\hat{k}_{1}^{2}+34}{6}\,e^{-6x_{0}}\,F(2,2,3,e^{-4x_{0}})
\nonumber\\
& &+\frac{16}{3}\,e^{-10x_{0}}\,F(3,3,4,e^{-4x_{0}})
\nonumber\\
& &+2\,e^{-14x_{0}}\, F(4,4,5,e^{-4x_{0}}).
\end{eqnarray}
It can be easily verified that
\begin{equation}
\mathcal{F}(-\hat{k}_{1},-\hat{k}_{1})=\mathcal{F}(\hat{k}_{1},\hat{k}_{1}),
\end{equation}
and
\begin{eqnarray}
\mathcal{F}(-\hat{k}_{1},\hat{k}_{1})&=&\frac{3\hat{k}_{1}^{2}-8i\hat{k}_{1}-
8}{3}\,\frac{1}{\sinh
2x_{0}}+\frac{3\hat{k}_{1}^{2}-12i\hat{k}_{1}-13}{6}
\,\frac{\cosh 2x_{0}}{\sinh^{2} 2x_{0}}
\nonumber\\
& 
&-\frac{2i\hat{k}_{1}+2}{3}\frac{\cosh^{2}2x_{0}}{\sinh^{3}2x_{0}}
-\frac{1}{6}
\,\frac{\cosh^{3}2x_{0}+5\cosh 2x_{0}}{\sinh^{4}2x_{0}}
\nonumber\\
& 
&-\frac{9\hat{k}_{1}^{3}-38i\hat{k}_{1}^{2}-50\hat{k}_{1}+20i}
{3(\hat{k}_{1}-2i)}
\,e^{-2x_{0}}\,F(1,\frac{i}{2}\hat{k}_{1}+1,\frac{i}{2}\hat{k}_{1}+2,
e^{-4x_{0}})
\nonumber\\
& 
&-\frac{6\hat{k}_{1}^{3}-62i\hat{k}_{1}^{2}-132\hat{k}_{1}+68i}
{3(\hat{k}_{1}-4i)}
\,e^{-6x_{0}}\,F(2,\frac{i}{2}\hat{k}_{1}+2,\frac{i}{2}\hat{k}_{1}+3,
e^{-4x_{0}})
\nonumber\\
& &+\frac{32i\hat{k}_{1}^{2}+136\hat{k}_{1}-96i}{3(\hat{k}_{1}-6i)}
\,e^{-10x_{0}}\,F(3,\frac{i}{2}\hat{k}_{1}+3,\frac{i}{2}\hat{k}_{1}+4,
e^{-4x_{0}})
\nonumber\\
& &+\frac{16\hat{k}_{1}-16i}{(\hat{k}_{1}-8i)}
\,e^{-14x_{0}}\,F(4,\frac{i}{2}\hat{k}_{1}+4,\frac{i}{2}\hat{k}_{1}+5,
e^{-4x_{0}}).
\end{eqnarray}
Finally $\mathcal{F}(\hat{k}_{1},-\hat{k}_{1})$ can be obtained from 
$\mathcal{F}(-\hat{k}_{1},\hat{k}_{1})$
after  setting $\hat{k}_{1}\rightarrow -\hat{k}_{1}$.

\resection{Type III(bulk-boundary)}

In this section we study the quantum correction to the classical
reflection 
factor due
to the contribution of the type III Feynman diagram, when  the
vertex associated with the loop is located at the bulk region and
the other vertex  coincides  with 
the boundary. The associated contribution is given by
\begin{eqnarray}\label{bulk-boundary}  
& &\mathcal{C}=-2\beta^{2}(\sigma_{1}\coth
x_{0}-\sigma_{0}\tanh x_{0})\int \int
\int 
dt dt' dx'
G(x_{1},t_{1};0,t)G(0,t;x',t')\nonumber\\
& &\hspace{2.1in}\times G(x',t';x',t')G(0,t;x_{2},t_{2})\sinh
(\sqrt{2}\beta \phi_{0}).
\end{eqnarray}

The following relation which is some part of the contribution
(\ref{bulk-boundary}), can
be derived independently from the remaining part
\begin{equation}
\mathcal{C}_{1}=\int dt G(x_{1},t_{1};0,t)G(0,t;x_{2},t_{2})  
\end{equation}
or 
\begin{eqnarray}
\mathcal{C}_{1}&= &\int dt \int\int
\frac{d\omega_{1}}{2\pi}\,\frac{dk_{1}}{2\pi}\,e^{-i\omega_{1}(t_{1}-t)}
\,\frac{i}{\omega_{1}^{2}-k_{1}^{2}-4+i\rho}\biggl(f(k_{1},x_{1})
f(-k_{1},0)
e^{ik_{1}x_{1}}\biggr.\nonumber\\
&
&\hspace{2.3in}\biggl.+K_{1}(k_{1})f(-k_{1},x_{1})f(-k_{1},0)
e^{-ik_{1}x_{1}}\biggr)
\nonumber\\
& &\times \int\int 
\frac{d\omega_{2}}{2\pi}\,\frac{dk_{2}}{2\pi}\,e^{-i\omega_{2}(t-t_{2})}
\,\frac{i}{\omega_{2}^{2}-k_{2}^{2}-4+i\rho}\biggl(f(k_{2},0)
f(-k_{2},x_{2})
e^{-ik_{2}x_{2}}\biggr.\nonumber\\
&
&\hspace{2.3in}\biggl.+K_{2}(k_{2})f(-k_{2},0)f(-k_{2},x_{2})
e^{-ik_{2}x_{2}}\biggr).
\end{eqnarray}
First of all, it is necessary to set $k_{1}\rightarrow -k_{1}$ in the
first term
of the first propagator.  Secondly, integration over $t$  leads to
the
substitution of 
$\omega_{2}=\omega_{1}$.  Finally integration over the momenta 
$k_{1}$ and
$k_{2}$, as before, may be done immediately by closing the contour 
in the upper
half-plane and looking at the poles at
$\hat{k}_{1}=k_{1}=k_{2}=\sqrt{\omega_{1}^{2}-4}$ and ignoring all 
the  other poles
 as their contributions  vanish rapidly as  $x_{1}, x_{2} \rightarrow
-\infty $. Therefore,  
\begin{eqnarray}\label{Green Green}
\mathcal{C}_{1}&=&\int\frac{d\omega_{1}}{2\pi}\,e^{-i\omega_{1}(t_{1}-t_{2})}
\,e^{-i\hat{k}_{1}(x_{1}+x_{2})}\,\frac{1}
{(2\hat{k}_{1})^{2}}A(\hat{k}_{1},x_{1})A(\hat{k}_{1},x_{2}),
\end{eqnarray}  
where
\begin{equation}
A(\hat{k}_{1},x_{1})=f(-\hat{k}_{1},x_{1})f(\hat{k}_{1},0)+
K(\hat{k}_{1})
f(-\hat{k}_{1},x_{1})f(-\hat{k}_{1},0).
\end{equation}
So, our next problem  is to calculate the following integral which is the
remaining part
of the contribution.
\begin{equation}\label{Green Green C}
\int\int dt'dx' G(0,t;x',t')G(x',t';x',t')\sinh(\sqrt{2}\beta 
\phi_{0}).
\end{equation}
Obviously, this part will be appeared as a constant
and it
must
be multiplied by  (\ref{Green Green}). Clearly, the time
variable
$t'$  appears only in one of the propagator i.e. in
$G(0,t;x',t')$. On
the other hand, this propagator along with integration over $t'$
has been
obtained in the previous section. Hence,
\begin{equation}
\int dt' G(0,t;x',t')=\frac{i}{8}e^{2x'}c_{0}\left(1-\coth
2(x'-x_{0})\right),
\end{equation}
where
\begin{equation}
c_{0}=-\left( 1+\tan^{2}\frac{(a_{0}+a_{1})\pi}{4}\right)
\left( 1+\tan^{2}\frac{(a_{0}-a_{1})\pi}{4}\right).
\end{equation}
 Therefore,  (\ref{Green Green C}) reduces to 
\begin{equation}\label{Green C}
\frac{ic_{0}}{8}\int dx' e^{2x'}\sinh(\sqrt{2}\beta \phi_{0}) 
\left(1-\coth
2(x'-x_{0})\right) G(x',t';x',t'),
\end{equation}
where
\begin{eqnarray}
G(x',t';x',t')= \int\int
\frac{d\omega'}{2\pi}\,\frac{dk'}{2\pi}
\,\frac{i}{\omega'^{2}-k'^{2}-4+i\rho}\biggl(f(k',x')f(-k',x')
\biggr.\nonumber\\
\biggl.+K(k')(f(-k',x')(f(-k',x')e^{-2ik'x'} \biggr)
\end{eqnarray}
or after integration over $\omega'$ 
\begin{eqnarray}
& &G(x',t';x',t')= \frac{1}{2}\int
\frac{dk'}{2\pi}
\,\frac{1}{\sqrt{k'^{2}+4}}\biggl(f(k',x')f(-k',x')
\biggr.\nonumber\\
& & \hspace{2in} \biggl.+K(k')(f(-k',x')(f(-k',x')e^{-2ik'x'}
\biggr).
\end{eqnarray}
In fact, the above integrand has two parts, 
the first part can be easily
manipulated
but the other part which includes the exponential term is hard to 
calculate and we
prefer to leave the computation  of that part for later. 
Let us look at the
first part
of the loop propagator. 
The integral of this part is logarithmically divergent.
Nevertheless, this divergence can be removed by an  infinite
renormalization of the mass parameter in the bulk potential.
Then, doing the integration over $k'$, we obtain

\begin{equation}  
 \frac{1}{2}\int
\frac{dk'}{2\pi}
\,\frac{1}{\sqrt{k'^{2}+4}}f(k',x')f(-k',x')=
-\frac{\left(1-\coth^{2}2(x'-x_{0})\right)}{2\pi}.
\end{equation}

To sum up, the integral (\ref{Green C})  reduces to 
\begin{eqnarray}
-\frac{ic_{0}}{16\pi}\int_{-\infty}^{0}dx'\,e^{2x'}\,\biggl(1-\coth^{2}
2(x'-x_{0})\biggr)
\biggl(1-\coth 2(x'-x_{0})\biggr)\sinh(\sqrt{2}\beta\phi_{0})
\nonumber\\
+\frac{ic_{0}}{16}\int_{-\infty}^{0}dx'
\int\frac{dk'}{2\pi}\,\frac{1}{\sqrt{k'^{2}+4}}
\biggl(1-\coth
2(x'-x_{0})\biggr)\sinh(\sqrt{2}\beta\phi_{0})\,e^{2x'}
\nonumber\\
\times 
\left\{\frac{(ik')^{2}+2ik'\sqrt{1+\sigma_{0}}\sqrt{1+\sigma_{1}}+
2(\sigma_{0}+\sigma_{1})}
{(ik')^{2}-2ik'\sqrt{1+\sigma_{0}}\sqrt{1+\sigma_{1}}+2(\sigma_{0}+
\sigma_{1})}
\frac{\left(ik'+2\coth
2(x'-x_{0})\right)^{2}}{(ik'+2)(ik'-2)}e^{-2ik'x'}\right\}
\end{eqnarray}
The above relation has two parts and the first part which is a 
single integral can
be performed by means of the formulae in  Appendix A  and we write
down only 
the
solution of
this part which is expressed in terms of hypergeometric functions, 
that is, 
\begin{eqnarray}
\mathcal{C}_{2}&= 
&-\frac{ic_{0}}{16\pi}\int_{-\infty}^{0}dx'e^{2x'}\left(1-\coth^{2}
2(x'-x_{0})\right)
\left(1-\coth 2(x'-x_{0})\right)\sinh(\sqrt{2}\beta\phi_{0})
\nonumber\\
&=&\frac{i}{16\pi}\left(1+\tan^{2}\frac{(a_{0}+a_{1})\pi}{4}\right)
\left(1+\tan^{2}\frac{(a_{0}-a_{1})\pi}{4}\right)
\nonumber\\
& & \,\,\,\times \left\{
\frac{1}{3\sinh 2x_{0}}-
\frac{\cosh 2x_{0}}{24\sinh^{2} 2x_{0}}
-\frac{\cosh^{2}2x_{0}+1}{6\sinh^{3}2x_{0}}-
\frac{\cosh^{3}2x_{0}+5\cosh 2x_{0}}{24\sinh^{4}2x_{0}}\right.
\nonumber\\
&
&\,\,\,\,\,\,\,\,\,\,\left.-\frac{1}{6}\,e^{-2x_{0}}\,F(1,1,2,
e^{-4x_{0}})\right.\nonumber\\
&
&\,\,\,\,\,\,\,\,\,\,\left.+\frac{11}{12}\,e^{-6x_{0}}\,F(2,2,3,
e^{-4x_{0}})\right.\nonumber\\
&
&\,\,\,\,\,\,\,\,\,\,\left.+\frac{4}{3}\,e^{-10x_{0}}\,F(3,3,4,
e^{-4x_{0}})\right.\nonumber\\
&
&\,\,\,\,\,\,\,\,\,\,\left.+\frac{1}{2}\,e^{-14x_{0}}\,F(4,4,5,
e^{-4x_{0}})\right\}.
\end{eqnarray}
So, in connection with the type III (bulk-boundary) contribution, the
remaining
integral is
\begin{eqnarray}\label{bulk-boundary 1}
\mathcal{C}_{3}=\frac{ic_{0}}{16}\int_{-\infty}^{0}dx'
\int\frac{dk'}{2\pi}\,\frac{1}{\sqrt{k'^{2}+4}}
\biggl(1-\coth 2(x'-x_{0})\biggr)\sinh(\sqrt{2} \beta \phi_{0}) 
\,e^{2x'}
\nonumber\\ 
\times 
\frac{\left(ik'+2\cos\frac{(a_{0}+a_{1})\pi}{2}\right)\left(ik'+
2\cos\frac{(a_{0}-a_{1}
)\pi}{2}\right)}{\left(ik'-2\cos\frac{(a_{0}+a_{1})\pi}{2}\right)
\left(ik'-2\cos
\frac{(a_{0}-a_{1})\pi}{2}\right)}\frac{\biggl(ik'+2\coth
2(x'-x_{0})\biggr)^{2}}{(ik'+2)(ik'-2)}e^{-2ik'x'}.
\end{eqnarray}
As we mentioned before, it is more convenient to integrate over $x'$
then
afterwards  over  $k'$. Since to integrate over $k'$ first is a
difficult problem.   
 Let us do partial fraction decomposition for the rational 
function in  (\ref{bulk-boundary 1}).  Obviously we 
will
have
four
elementary partial fraction
including 
\par
$$\frac{1}{\left(ik'-2\cos\frac{(a_{0}+a_{1})\pi}{2}\right)},
\ \frac{1}{\left(ik'-2\cos\frac{(a_{0}-a_{1})\pi}{2}\right)},
\ \frac{1}{ik'+2},\ \frac{1}{ik'-2}.$$ 

Now, in what follows we perform the calculations in detail for one
of
them, for
example, 
$\frac{1}{\left(ik'-2\cos\frac{(a_{0}+a_{1})\pi}{2}\right)}$ 
due to the 
fact that for all the others the computations are similar except
that $\cos\frac{(a_{0}+a_{1})\pi}{2}$ is replaced by one of 
$\cos\frac{(a_{0}-a_{1})\pi}{2}$, -1, 1, respectively. What we need
to do is to calculate the following :
\begin{eqnarray}\label{bulk-boundary 2}
& &
-\frac{\left(\tan ^{2}\frac{(a_{0}+a_{1})\pi}{4}
-\cot^{2}\frac{(a_{0}+a_{1})\pi}{4}\right)}
{\cos^{2}\frac{(a_{0}+a_{1})\pi}{4}\cos^{2}
\frac{(a_{0}-a_{1})\pi}{4}}
\,\cot \frac{a_{0}\pi}{2}\,\cot \frac{a_{1}\pi}{2}
\nonumber\\
& &\times\, \frac{i}{16}\int_{-\infty}^{0}dx'
\int\frac{dk'}{2\pi}\,\frac{1}{\sqrt{k'^{2}+4}}
\,\biggl(1-\coth 2(x'-x_{0})\biggr)\sinh(\sqrt{2} \beta \phi_{0})
\,e^{(2-2ik')x'}
\nonumber\\
& &\hspace{.5in}\times\, \left(\coth
2(x'-x_{0})
+\cos \frac{(a_{0}+a_{1})\pi}{2}\right)^{2}
\left(\frac{1}{ik'-2\cos\frac{(a_{0}+a_{1})\pi}{2}}\right).
\end{eqnarray}      
The integration  over $x'$ may be done by using the formulae in Appendix
A and gives 
\begin{eqnarray}
& &\int_{-\infty}^{0}dx'\,e^{(2-2ik')x'}\,\biggl(1-\coth
2(x'-x_{0})\biggr) \sinh(\sqrt{2} \beta \phi_{0}) 
\nonumber\\
& &\hspace{1.1in}\times\,\left(\coth    
2(x'-x_{0})                                                 
+\cos \frac{(a_{0}+a_{1})\pi}{2}\right)^{2}
\nonumber\\
& =& L(a_{0},a_{1})
\nonumber\\
&+&\sum_{n=1}^{4}\left\{\frac{(A_{n} k'+B_{n})}{(k'+2ni)}
\,e^{-(2+4(n-1))x_{0}}\, 
F(n,-\frac{i}{2}k'+n,-\frac{i}{2}k'+n+1,e^{-4x_{0}})\right\},
\end{eqnarray}  
where
\begin{eqnarray}
L(a_{0},a_{1}) &=& \left(\cos^{2}\frac{(a_{0}+a_{1})\pi}{2}
-\frac{4}{3}\cos\frac{(a_{0}+a_{1})\pi}{2}+\frac{2}{3}\right)
\frac{1}
{\sinh 2x_{0}}
\nonumber\\
&
+&\left(\frac{1}{2}\cos^{2}\frac{(a_{0}+a_{1})\pi}{2}-\cos
\frac{(a_{0}+a_{1})\pi}{2}
+\frac{13}{24}\right)\frac{\cosh 2x_{0}}{\sinh^{2}2x_{0}}
\nonumber\\
&
-&\left(\frac{1}{3}\cos\frac{(a_{0}+a_{1})\pi}{2}-
\frac{1}{6}\right)
\frac{\cosh^{2}2x_{0}+1}{\sinh^{3}2x_{0}}
\nonumber\\
& +&\frac{1}{24}\,\frac{\cosh^{3}2x_{0}
+5\cosh 2x_{0}}{\sinh^{4}
2x_{0}}
\end{eqnarray}
and  the coefficients $A_{n}$, $B_{n}$, $n=1, 2, 3, 4$ are 
constants which in fact 
  only  depend on $\cos\frac{(a_{0} + a_{1})\pi}{2}$. 
Now the final
calculation is to integrate
over $k'$ and it is evident that in order to do this, 
we have to convert
the
hypergeometric functions to infinite series. Considering the
equation (\ref{hypergeometric 1})  in
Appendix A, we  conclude  that 
\begin{equation}
\frac{ F(1,-\frac{i}{2}k'+1,-\frac{i}{2}k'+2,e^{-4x_{0}})}{k'+2i}=
\sum_{n=0}^{\infty}\frac{e^{-4nx_{0}}}{k'+i(2+2n)}.
\end{equation}  
If we differentiate    both sides of the above relation with 
respect to
$x_{0}$,
then the following identity may be derived

\begin{equation}
\frac{ F(2,-\frac{i}{2}k'+2,-\frac{i}{2}k'+3,e^{-4x_{0}})}{k'+4i}=
\sum_{n=1}^{\infty}\frac{ne^{-4(n-1)x_{0}}}{k'+i(2+2n)}.
\end{equation}  
In the same way, one obtains
\begin{equation}
\frac{ F(3,-\frac{i}{2}k'+3,-\frac{i}{2}k'+4,e^{-4x_{0}})}{k'+6i}=
\frac{1}{2!}\sum_{n=2}^{\infty}
\frac{n(n-1)e^{-4(n-2)x_{0}}}{k'+i(2+2n)}
\end{equation}
and
\begin{equation}
\frac{ F(4,-\frac{i}{2}k'+4,-\frac{i}{2}k'+5,e^{-4x_{0}})}{k'+8i}=
\frac{1}{3!}\sum_{n=3}^{\infty}\frac{n(n-1)(n-2)
e^{-4(n-3)x_{0}}}{k'+i(2+2n)}.
\end{equation}

Now if we substitute  (6.19), (6.20), (6.21) 
and (6.22) into  (6.17),
all that remains in connection with the contribution  (6.16)
  is the integration over  $k'$. Obviously we encounter 
 integrals of the form  
\begin{equation}
\int_{-\infty}^{\infty}\frac{dk'}{\sqrt{k'^{2}+4}}\,
\left(\frac{1}{ik'-2
\cos\frac{(a_{0}+a_{1})\pi}{2}}\right)\,
\left(\frac{Ak'+B}{k'+i(2+2n)}\right)
\end{equation} 
and the $k'$ integration may be 
performed by closing the contour in the upper
half-plane and onto the 
branch cut which  stretches from $k'=2i$ to 
infinity
along the imaginary axis.  In fact, leaving the integrals along the
branch
cut  to be evaluated later,   we obtain the required formula 
\begin{eqnarray}\label{ln1}
& &\int_{-\infty}^{\infty}\frac{dk'}{\sqrt{k'^{2}+4}}\, 
\left(\frac{1}{ik'-2
\cos\frac{(a_{0}+a_{1})\pi}{2}}\right)\,
\left(\frac{Ak'+B}{k'+i(2+2n)}\right)
\nonumber\\
& 
&=\frac{\left(2\cos\frac{(a_{0}+a_{1})\pi}{2}A+iB\right)}
{\left(2n+2-2\cos\frac{(a_{0}+a_{1})\pi}{2}\right)}
\, \frac{\frac{(a_{0}+a_{1})\pi}{2}}{\sin\frac{(a_{0}+a_{1})\pi}{2}}
\nonumber\\
& 
&\,\,\,\,\,\,-\frac{\left((2n+2)A+iB\right)}{\left(2n+2-2\cos
\frac{(a_{0}+a_{1})\pi}{2}\right)}
\,
\frac{1}{2\sqrt{n^{2}+2n}}\ln\left\{{\frac{n+1-\sqrt{n^{2}+2n}}
{n+1+\sqrt{n^{2}+2n}}}\right\}.
\end{eqnarray}
Note  (\ref{ln1})  is valid 
when $n\neq 0$, on the other hand if $n=0$ then
one may find
\begin{eqnarray}
& &\int_{-\infty}^{\infty}\frac{dk'}{\sqrt{k'^{2}+4}}\, 
\left(\frac{1}{ik'-2
\cos\frac{(a_{0}+a_{1})\pi}{2}}\right)\,
\left(\frac{Ak'+B}{k'+2i}\right)
\nonumber\\
& 
&=\frac{\left(2\cos
\frac{(a_{0}+a_{1})\pi}{2}A+iB\right)}{\left(2-2\cos
\frac{(a_{0}+a_{1})\pi}{2}\right)}
\, \frac{\frac{(a_{0}+a_{1})\pi}{2}}{\sin\frac{(a_{0}+a_{1})\pi}{2}}
-\frac{\left(2A+iB\right)}{\left(2-2\cos\frac{(a_{0}+a_{1})\pi}{2}\right)}.
\end{eqnarray}

Now we are in a position to write down 
(\ref{bulk-boundary 2})
or, in fact, the integral (6.15) 
\begin{eqnarray}\label{bulk-boundary 3}
\mathcal{C}_{3}&=&\frac{i}{32\pi}\,
\frac{\left(\tan ^{2}\frac{(a_{0}+a_{1})\pi}{4}
-\cot^{2}\frac{(a_{0}+a_{1})\pi}{4}\right)}
{\cos^{2}\frac{(a_{0}+a_{1})\pi}{4}\cos^{2}
\frac{(a_{0}-a_{1})\pi}{4}}
\,\cot \frac{a_{0}\pi}{2}\,\cot \frac{a_{1}\pi}{2}
\nonumber\\
& &\times \left\{ 
  \frac{\frac{(a_{0}+a_{1})\pi}{2}}
{\sin\frac{(a_{0}+a_{1})\pi}{2}}\, L(a_{0},a_{1}) 
\right.\nonumber\\
& &\left.\,\,\,\,+\frac{1}{12}\,e^{-2x_{0}}\left(
\frac{2A_{1}+iB_{1}}{2-2\cos\frac{(a_{0}+a_{1})\pi}{2}} 
-\frac{2\cos\frac{(a_{0}+a_{1})\pi}{2}A_{1}+iB_{1}}
{2-2\cos\frac{(a_{0}+a_{1})\pi}{2}}
\,\frac{\frac{(a_{0}+a_{1})\pi}{2}}
{\sin\frac{(a_{0}+a_{1})\pi}{2}}\right)
\right. \nonumber\\
& &\left.\,\,\,\,-\sum_{n=1}^{\infty}
\frac{\frac{(a_{0}+a_{1})\pi}{2}}{\sin\frac{(a_{0}+a_{1})\pi}{2}}
\,\frac{e^{-(2+4n)x_{0}}}{\left(2n+2-2\cos
\frac{(a_{0}+a_{1})\pi}{2}\right)}
\Biggl[
\left(2\cos\frac{(a_{0}+a_{1})\pi}{2}A_{1}+iB_{1}\right)
\Biggr.\right.\nonumber\\
&
&\Biggl.\left.\hspace{.24in}
+\frac{n}{1!}\left(2\cos\frac{(a_{0}+a_{1})\pi}{2}A_{2}+iB_{2}
\right)
+\frac{n(n-1)}{2!}\left(2\cos\frac{(a_{0}+a_{1})\pi}{2}A_{3}+iB_{3}
\right)
\Biggr.\right.\nonumber\\
& &\hspace{2.1in} 
\Biggl.\left.+\frac{n(n-1)(n-2)}{3!}
\left(2\cos\frac{(a_{0}+a_{1})\pi}{2}A_{4}+i
B_{4} \right) \Biggr] 
\right.\nonumber\\
&
&\left.\,\,\,\,+
\sum_{n=1}^{\infty}
\frac{e^{-(2+4n)x_{0}}}{\left(2n+2-2\cos\frac{(a_{0}+a_{1})\pi}
{2}\right)}\,\frac{1}{2\sqrt{n^{2}+2n}}
\ln{\left\{\frac{n+1-\sqrt{n^{2}+2n}}{n+1+\sqrt{n^{2}+2n}}\right\}}
\right.\nonumber\\
& & \hspace{1.5in} 
\left.\biggl(\biggl[ (2n+2)A_{1}+iB_{1}\biggr]
+\frac{n}{1!}\biggl[  (2n+2)A_{2}+iB_{2}\biggr]
\biggr.\right.\nonumber\\
& &\hspace{.4in}\biggl.\left.+\frac{n(n-1)}{2!}\biggl[
(2n+2)A_{3}+iB_{3}\biggr]
+\frac{n(n-1)(n-2)}{3!}
\biggl[  (2n+2)A_{4}+iB_{4}\biggr]\biggr) \right\}  
\nonumber\\
& &\hspace{.1in}+ \,\,\,\hbox{other pole pieces}.
\end{eqnarray}
Note, in the above expression  all the series are convergent. As we
mentioned before,  
(\ref{bulk-boundary 3})  must be considered
(after adding to (6.14)) as a
coefficient factor of 
 (\ref{Green Green}) in order to constitute the
type III
(bulk-boundary) contribution i.e.:
\begin{equation}
\mathcal{C}=\mathcal{C}_{1}\left(\mathcal{C}_{2}+\mathcal{C}_{3}
 \right).
\end{equation}

\resection{Type I diagram}

In this section we calculate 
the contribution of the type I Feynman diagram to the classical
reflection factor when the vertex is placed inside the bulk region.
In fact, when the vertex is located at the boundary then, the
corresponding contribution has been found \cite{MA} and    is
given by
\begin{eqnarray}
& &-\frac{i\beta^{2}}{8}   
(\sigma_{1}\coth x_{0}+\sigma_{0}\tanh
x_{0})(
a_{0}\cot a_{0}\pi+a_{1}\cot a_{1}\pi)
\nonumber\\
& &\times \int\frac{d\omega}{2\pi}\,e^{-i\omega (t_{1}-t_{2})}
\,e^{-i\hat{k}(x_{1}+x_{2})}
\,\frac{i\hat{k}+2\coth
2(x_{1}-x_{0})}{P(\hat{k})}\,\frac{i\hat{k}+2\coth
2(x_{2}-x_{0})}{P(\hat{k})}
\end{eqnarray}

Clearly, in our case  the bulk four point
coupling    should be considered
 in the interaction
vertex. Moreover, as before, 
the combinatorial factor associated with this diagram will 
appear  as a coefficient. Hence, the contribution has the form

\begin{equation} \label{type I bulk}
-4i\beta^{2}\int_{-\infty}^{\infty}dt 
\int_{-\infty}^{0}dx G(x_{1},t_{1};x,t)G(x,t;x,t)
G(x,t;x_{2},t_{2})\cosh(\sqrt{2}\beta\phi_{0}),
\end{equation}     
where
\begin{equation}
\cosh(\sqrt{2}\beta\phi_{0})=\left( 2\coth^{2}2(x-x_{0})-1\right).
\end{equation}

In the previous section, 
we simplified the loop propagator
to  
\begin{eqnarray}
G(x,t;x,t)&=&-\frac{\left(1-\coth^{2}2(x-x_{0})\right)}{2\pi}
\nonumber\\
& &+\frac{1}{2}\int
\frac{dk''}{2\pi}\,\frac{1}{\sqrt{k''^{2}+4}}
\,K(k'')f(-k'',x)f(-k'',x)\,e^{-2ik''x}.
\end{eqnarray}
Also the integral part of the loop Green function is hard enough
to evaluate  and 
we found out that it is better to do this integration during
the final stage. Now, let us 
rewrite the   contribution
 (\ref{type I bulk}) 
in the 
expanded
form 
\begin{eqnarray}\label{bulk 0}
& &-4i\beta^{2}\int dt 
\int_{-\infty}^{0} dx\int \int
\frac{d\omega}{2\pi}\,\frac{dk}{2\pi}\,e^{-i\omega(t_{1}-t)}\,
\frac{i}{\omega^{2}-k^{2}-4+i\rho} \biggl(f(k,x_{1})f(-k,x)
e^{ik(x_{1}-x)}
\biggr. \nonumber\\
& &\hspace{2in}\biggl.+K(k)f(-k,x_{1})f(-k,x)e^{-ik(x_{1}+x)}
\biggr)\cosh(\sqrt{2}\beta\phi_{0})
\nonumber\\
& &\times \left\{-\frac{\left(1-\coth^{2}2(x-x_{0})\right)}{2\pi}
+\frac{1}{2}\int
\frac{dk''}{2\pi}\frac{1}{\sqrt{k''^{2}+4}}
K(k'')f(-k'',x)f(-k'',x)e^{-2ik''x}
\right\}\nonumber\\
& &\times \int \int
\frac{d\omega'}{2\pi}\,\frac{dk'}{2\pi}\,e^{-i\omega'(t-t_{2})}\,
\frac{i}{\omega'^{2}-k'^{2}-4+i\rho} \left(f(k',x)f(-k',x_{2})
e^{ik'(x-x_{2})}
\right. \nonumber\\  
& &\hspace{2.5in}\left.+K'(k')f(-k',x)f(-k',x_{2})e^{-ik'(x+x_{2})}
\right).
\end{eqnarray}

Looking at  (\ref{bulk 0}), 
one can  predict that the calculations will be 
lengthy
and intricate. 
 The starting point is to do the $t$ integration which
allows 
 the
substitution  $\omega=\omega'$. Secondly, 
it is necessary to perform a
transformation 
$k \rightarrow -k $ in the 
first term of the first propagator. Moreover, if we
multiply the
first and the third 
propagator with  each other, then obviously we will have  four
pole pieces
and fortunately if we do the 
calculation for one of them (for example the first one), then the
calculations corresponding to the 
other three pole pieces may be treated similarly with  $k+k'$ 
replaced by one of $k-k'$, $-k+k'$ and $-k-k'$.
Because of this  
in what follows we follow the problem only for
one pole piece. Hence
our problem  is, in fact, the following integral
\begin{eqnarray}
& &\mathcal{D}=-4i\beta^{2}
\int_{-\infty}^{0} dx\int \int
\frac{d\omega}{2\pi}\,\frac{dk}{2\pi}\,e^{-i\omega(t_{1}-t_{2})}\,
\frac{i}{\omega^{2}-k^{2}-4+i\rho} 
\nonumber\\
&
&\nonumber\\
& &
\hspace{1in}\times 
\cosh (\sqrt{2}\beta \phi_{0})f(-k,x_{1})f(k,x)
\,e^{-ik(x_{1}-x)}
\nonumber\\
& &\nonumber\\
& &\,\,\times \left(-\frac{\left(1-\coth^{2}2(x-x_{0})\right)}{2\pi}
+\frac{1}{2}\int
\frac{dk''}{2\pi}\,\frac{1}{\sqrt{k''^{2}+4}}
K(k'')f(-k'',x)f(-k'',x)e^{-2ik''x}\right)
\nonumber\\
& &\,\,\times  \int\frac{dk'}{2\pi}
\,\frac{i}{\omega^{2}-k'^{2}-4+i\rho} f(k',x)f(-k',x_{2})
\,e^{ik'(x-x_{2})}.
\end{eqnarray}
In fact, the above contribution has two parts. The first part,  in
which
the integral of the middle momentum $(k'')$ is not involved,  can be
calculated by means
of  the formulae in   Appendix B  and we call this part
$\mathcal{D}_{1}$. 
Let us  write down
the solution of this part.
This contribution is expressed in terms 
of the hypergeometric function
as :
\begin{eqnarray}\label{bulk 1}
\mathcal{D}_{1}&=&\frac{i\beta^{2}}{\pi} \int
\frac{d\omega}{2\pi}\,e^{-i\omega(t_{1}-t_{2})}
\,e^{-i\hat{k}(x_{1}+x_{2})}f(-\hat{k},x_{1})f(-\hat{k},x_{2})
\frac{1}{\hat{k}^{2}}
\nonumber\\
& &
\hspace{.5in}\times \left\{\frac{4i-4\hat{k}-i\hat{k}^{2}}
{\hat{k}-2i}
\, e^{-4x_{0}}
\, 
F\left(2,\frac{i}{2}\hat{k}+1,\frac{i}{2}\hat{k}+2,e^{-4x_{0}}\right)
\right. \nonumber\\
& & \hspace{.8in} 
\left.+\frac{48i-40\hat{k}-8i\hat{k}^{2}}{\hat{k}-4i}
\, e^{-8x_{0}}
\, 
F\left(3,\frac{i}{2}\hat{k}+2,\frac{i}{2}\hat{k}+3,e^{-4x_{0}}\right)
\right. \nonumber\\
& & \hspace{.8in} 
\left.+\frac{176i-96\hat{k}-8i\hat{k}^{2}}{\hat{k}-6i}
\, e^{-12x_{0}}
\, 
F\left(4,\frac{i}{2}\hat{k}+3,\frac{i}{2}\hat{k}+4,e^{-4x_{0}}\right)
\right. \nonumber\\
& & \hspace{.8in} \left.+\frac{256i-64\hat{k}}{\hat{k}-8i}
\, e^{-16x_{0}}
\, 
F\left(5,\frac{i}{2}\hat{k}+4,\frac{i}{2}\hat{k}+5,e^{-4x_{0}}\right)
\right. \nonumber\\
& &\hspace{.8in}  \left.+\frac{128i}{\hat{k}-10i}
\, e^{-20x_{0}}
\, 
F\left(6,\frac{i}{2}\hat{k}+5,\frac{i}{2}\hat{k}+6,e^{-4x_{0}}\right)
\right\}.
\end{eqnarray}

Now it is better for the second part,  which we call  
$\mathcal{D}_{2}$,   to
integrate first over
$x$ then over $k''$.
Meanwhile, before
starting the integration, it is useful to 
note that if we do the partial fraction decomposition for
  $K''(k'')f(-k'',x)f(-k'',x)$, 
then we will have four elementary partial fractions as
\par
$$\frac{1}{\left(ik''-2\cos\frac{(a_{0}+a_{1})\pi}{2}\right)},
\ \frac{1}{\left(ik''-2\cos\frac{(a_{0}-a_{1})\pi}{2}\right)},
\ \frac{1}{ik''+2},\ \frac{1}{ik''-2}.$$
As before, in the remaining section 
we continue the computations in detail for one of them (for
example,  $\frac{1}{\left(ik''-2\cos
\frac{(a_{0}+a_{1})\pi}{2}\right)}$) because the
calculations corresponding to the 
other three elementary partial fractions can be done in the
same manner just by the substitution of
$\cos\frac{(a_{0}+a_{1})\pi}{2}$ by one of  
$\cos\frac{(a_{0}-a_{1})\pi}{2}$, -1, 1, respectively.
So, our problem reduces to this integral
\begin{eqnarray}\label{bulk 2}
& &2i\beta^{2}
\,\cot \frac{a_{0}\pi}{2}\,\cot \frac{a_{1}\pi}{2}
\left(\tan ^{2}\frac{(a_{0}+a_{1})\pi}{4}
-\cot^{2}\frac{(a_{0}+a_{1})\pi}{4}\right)
\nonumber\\
& &\times\, \int_{-\infty}^{0} dx\int \int
\frac{d\omega}{2\pi}\,\frac{dk}{2\pi}\,e^{-i\omega(t_{1}-t_{2})}\,
\frac{i}{\omega^{2}-k^{2}-4+i\rho} 
\nonumber\\
& &\nonumber\\
& &\hspace{1in}\times\, 
\cosh (\sqrt{2}\beta \phi_{0})f(-k,x_{1})f(k,x) \,e^{ik(x_{1}-x)}
\nonumber\\
& &\nonumber\\
& &
\times \int
\frac{dk''}{2\pi}\,\frac{1}{\sqrt{k''^{2}+4}}\,\frac{e^{-2ik''x} 
}{ik''-2\cos\frac{(a_{0}+a_{1})\pi}{2}}\,
\left(\coth 2(x-x_{0})+\cos\frac{(a_{0}+a_{1})\pi}{2}\right)^{2}   
\nonumber\\
& & \times\, \int\frac{dk'}{2\pi}
\,\frac{i}{\omega^{2}-k'^{2}-4+i\rho} f(k',x)f(-k',x_{2})
\,e^{ik'(x-x_{2})}.
\end{eqnarray}
The integration over  $x$ can be evaluated by 
means of the formulae given in Appendix B as 
\begin{eqnarray}
& &\int_{-\infty}^{0}dx \,e^{i(k+k'-2k'')x}
\cosh(\sqrt{2}\beta\phi_{0})f(k,x)f(k',x)
 \left(\coth
2(x-x_{0})+\cos\frac{(a_{0}+a_{1})\pi}{2}\right)^{2}
\nonumber\\
&
&=\frac{1}{(ik+2)(ik'+2)}
\frac{\left(A'_{0}kk'+B'_{0}(k+k')+C'_{0}\right)}{(k+k'-2k'')}
\nonumber\\
& &+\frac{1}{(ik+2)(ik'+2)}
\sum_{n=1}^{6}\left\{
\frac{\left(A'_{n}kk'+B'_{n}(k+k')+C'_{n}\right)}{\left(k+k'-2k''-4(n-1)i
\right)}
\,e^{-4(n-1)x_{0}}\,
\right.\nonumber\\
& 
&\hspace{.45in}\left. 
\times 
\, 
F\left(n,\frac{i}{4}(k+k'-2k'')+n-1,\frac{i}{4}(k+k'-2k'')+n,e^{-4x_{0}}
\right)
\right\},
\end{eqnarray}
where the coefficients $A'_{n}, B'_{n}, C'_{n}; n=0,1,..,6 $ are
constants and  depend  only on  
$\cos\frac{(a_{0} + a_{1})\pi}{2}$. In fact $A'_{5}$, $A'_{6}$
and $B'_{6}$ are zero. Now the subsequent calculation is
to integrate over
$k''$ and it is clear that to do  this, 
it is necessary to convert the
hypergeometric
function to an  infinite series. 
Looking at  (\ref{hypergeometric 2}), we may write down
\begin{equation}
\frac{F(1,\frac{i}{4}(k+k'-2k''),
\frac{i}{4}(k+k'-2k'')+1,e^{-4x_{0}})}{k+k'-2k''}
=\sum_{n=0}^{\infty}\frac{e^{-4nx_{0}}}{k+k'-2k''-4ni}
\end{equation}
and by differentiating    both sides of the above relation with
respect to $x_{0}$,
then
we obtain 
\begin{equation}
\frac{F(2,\frac{i}{4}(k+k'-2k'')+1,
\frac{i}{4}(k+k'-2k'')+2,e^{-4x_{0}})}{k+k'-2k''-4i}
=\sum_{n=1}^{\infty}\frac{n e^{-4(n-1)x_{0}}}{k+k'-2k''-4ni}. 
\end{equation}  
Similarly one can derive the 
infinite series forms  of the other hypergeometric functions as 
\begin{equation}
\frac{F(3,\frac{i}{4}(k+k'-2k'')+2,
\frac{i}{4}(k+k'-2k'')+3,e^{-4x_{0}})}{k+k'-2k''-8i}
=\frac{1}{2!}
\sum_{n=2}^{\infty}\frac{n(n-1) e^{-4(n-2)x_{0}}}{k+k'-2k''-4ni},
\end{equation}
\begin{eqnarray}
& &\frac{F(4,\frac{i}{4}(k+k'-2k'')+3,
\frac{i}{4}(k+k'-2k'')+4,e^{-4x_{0}})}{k+k'-2k''-12i}
\nonumber\\
& &\hspace{1in}=\frac{1}{3!}\sum_{n=3}^{\infty}\frac{n(n-1)(n-2)
e^{-4(n-3)x_{0}}}{k+k'-2k''-4ni},
\end{eqnarray}
\begin{eqnarray}
& &\frac{F(5,\frac{i}{4}(k+k'-2k'')+4,
\frac{i}{4}(k+k'-2k'')+5,e^{-4x_{0}})}{k+k'-2k''-16i}
\nonumber\\
& &\hspace{1in}=\frac{1}{4!}\sum_{n=4}^{\infty}\frac{n(n-1)(n-2)(n-3)
e^{-4(n-4)x_{0}}}{k+k'-2k''-4ni}
\end{eqnarray}   
and
\begin{eqnarray}
& &\frac{F(6,\frac{i}{4}(k+k'-2k'')+5,
\frac{i}{4}(k+k'-2k'')+6,e^{-4x_{0}})}{k+k'-2k''-20i}
\nonumber\\
& &\hspace{1in}=\frac{1}{5!}\sum_{n=5}^{\infty}\frac{n(n-1)(n-2)(n-3)(n-4)
e^{-4(n-5)x_{0}}}{k+k'-2k''-4ni}.
\end{eqnarray}

Let us substitute  (7.10), (7.11), 
(7.12), (7.13),  (7.14) and (7.15) in 
(7.9) 
and
obviously what  remains 
in connection with the contribution (7.8), 
   are the  integrations over 
$k''$, $k'$ and $k$ . As before in
previous sections, 
in order to integrate 
over the momenta $k$ and $k'$, it is sufficient to
close the contours in the upper half-plane and pick up poles at
$\hat{k}=k=k'=\sqrt{\omega^{2}-4}$ 
as  all the other poles'
contributions will be exponentially damped when 
$ x,x'\rightarrow -\infty $ . Meanwhile, the 
integration over $k''$  is of the form  

\begin{equation}
\int\frac{dk''}{\sqrt{k''^{2}+4}}
\,\frac{1}{\left(ik''-2\cos\frac{(a_{0}+a_{1})\pi}{2}\right)}
\,\frac{1}{\left(k+k'-2k''-4ni\right)}.
\end{equation}
Now to manipulate the  integral, 
let us choose the contour in the upper half-plane,
 taking care of   the branch cut which  runs  from $ k''=2i $ to
infinity
along the imaginary axis. 
Clearly this integral reduces to the integrals along the branch cut
and  we obtain
\begin{eqnarray}
& &\int\frac{dk''}{\sqrt{k''^{2}+4}}
\,\frac{1}{\left(ik''-2\cos\frac{(a_{0}+a_{1})\pi}{2}\right)}
\,\frac{1}{\left(k+k'-2k''-4ni\right)}
\nonumber\\
& & =\frac{1}{\left(k+k'+4i
\cos\frac{(a_{0}+a_{1})\pi}{2}-4ni\right)}
\left(-\frac{\frac{(a_{0}+a_{1})\pi}{2}}
{\sin\frac{(a_{0}+a_{1})\pi}{2}}
\right.\nonumber\\
&
&\,\,\,\,\,\,\,\,\,\,\left.-\frac{2i}
{\sqrt{\frac{(k+k')^{2}}{4}+4-4n^{2}-2ni(k+k')}}
\right.\nonumber\\
&
&\,\,\,\,\,\,\,\,\,\,\,\,\,\,\left.\times 
\ln\left\{{\frac{1+\frac{i(k+k')}{4}+n+
\frac{i}{2}\sqrt{\frac{(k+k')^{2}}{4}+4-4n^{2}-2ni(k+k')}}
{1+\frac{i(k+k')}{4}+n-\frac{i}{2}
\sqrt{\frac{(k+k')^{2}}{4}+4-4n^{2}-2ni(k+k')}}}\right\}\right).
\end{eqnarray}
When $ n=0 $, the above formula is simplified much more, especially
after doing
the integration over
$k$ and $k'$ and using the  fact that 
$ \hat{k}=k=k'=2\sinh \theta $ . So the following formula
can be obtained
\begin{eqnarray}
& &\int\frac{dk''}{\sqrt{k''^{2}+4}}
\,\frac{1}{\left(ik''-2\cos\frac{(a_{0}+a_{1})\pi}{2}\right)}
\,\frac{1}{\left(2\hat{k}-2k''\right)}
\nonumber\\
& &
\,\,\,\,=\frac{1}{\left(2\hat{k}+4i\cos\frac{(a_{0}+a_{1})\pi}{2}\right)}
\left(-\frac{\frac{(a_{0}+a_{1})\pi}{2}}
{\sin\frac{(a_{0}+a_{1})\pi}{2}}+
\frac{2}{\sqrt{\hat{k}^{2}+4}}(\frac{\pi}{2}-i\theta)
\right).
\end{eqnarray}
Now,  (7.8)  or more
generally the contribution $\mathcal{D}_{2}$  can be written
as:
 \begin{eqnarray}
\mathcal{D}_{2}&=
&\frac{i\beta^{2}}{4\pi}
\,\cot \frac{a_{0}\pi}{2}\,\cot \frac{a_{1}\pi}{2}
\left(\tan ^{2}\frac{(a_{0}+a_{1})\pi}{4}
-\cot^{2}\frac{(a_{0}+a_{1})\pi}{4}\right)
\nonumber\\
& & \times \int
\frac{d\omega}{2\pi}\,e^{-i\omega(t_{1}-t_{2})}
\,e^{-i\hat{k}(x_{1}+x_{2})}\,
\left(\frac{1}{\hat{k}}\right)^{2}f(-\hat{k},x_{1})f(-\hat{k},x_{2})
\nonumber\\
&
&\times 
\left\{\frac{2i\left(1-\cos
\frac{(a_{0}+a_{1})\pi}{2}\right)^{2}}{\left(\hat{k}+2i\cos\frac
{(a_{0}+a_{1})\pi}{2}\right)}\left(\frac{\frac{(a_{0}+a_{1})\pi}{2}}
{\sin\frac{(a_{0}+a_{1})\pi}{2}}-
\frac{2}{\sqrt{\hat{k}^{2}+4}}(\frac{\pi}{2}-i\theta)
\right)
\right.\nonumber\\
& &\hspace{.2in}\left.-\frac{1}{(i\hat{k}+2)^{2}}
\sum_{n=1}^{\infty}\frac{e^{-4nx_{0}}}{\left(2\hat{k}+4i
\cos\frac{(a_{0}+a_{1})\pi}{2}-4ni\right)}
\left(\frac{\frac{(a_{0}+a_{1})\pi}{2}}
{\sin\frac{(a_{0}+a_{1})\pi}{2}}
\right.\right.\nonumber\\
&
&\left.\left.\hspace{.5in}+
\frac{2i}{\sqrt{\hat{k}^{2}+4-4n^{2}-4ni\hat{k}}}
\ln{\left\{\frac{1+\frac{i}{2}\hat{k}+n+\frac{i}{2}\sqrt{\hat{k}^{2}
+4-4n^{2}-4ni\hat{k}}}
{1+\frac{i}{2}\hat{k}+n-\frac{i}{2}\sqrt{\hat{k}^{2}+4-4n^{2}-
4ni\hat{k}}}\right\}}
\right)\right. \nonumber\\
& & \left.\left( (A'_{1}\hat{k}^{2}+2B'_{1}\hat{k}+C'_{1})
+n(A'_{2}\hat{k}^{2}+2B'_{2}\hat{k}+C'_{2})
+\frac{n(n-1)}{2!}(A'_{3}\hat{k}^{2}+2B'_{3}\hat{k}+C'_{3})
\right. \right.\nonumber\\
& &\left.\left.
+\frac{n(n-1)(n-2)}{3!}(A'_{4}\hat{k}^{2}+2B'_{4}\hat{k}+C'_{4})
+\frac{n(n-1)(n-2)(n-3)}{4!}(2B'_{5}\hat{k}+C'_{5})
\right.\right.\nonumber\\
& &\left.\left.+\frac{n(n-1)(n-2)(n-3)(n-4)}{5!}C'_{6}\right) 
\right\}
\nonumber\\
& &+\,\, \hbox{other pole pieces}.
\end{eqnarray}
Firstly, in order to check the above solution, if we set
$a_{0}=a_{1}$ and consider the other pole pieces then, we can
derive the formula (3.10) in reference \cite{C}. As we mentioned
before, the calculation of this reference is based on the case when
the boundary parameters are equal. Secondly, in this solution, we
verified that  the term which depends  explicitly on the rapidity
of the particle ($\theta$) is cancelled by counterpart terms in the
other pole pieces. It is evident that if we add the expressions 
(7.7) and
(7.19)
then, the contribution (7.6)  will be obtained i.e.
$\mathcal{D}=\mathcal{D}_{1}+\mathcal{D}_{2}$.

\resection{Discussion}
Affine Toda field theory on 
the whole line is an exactly solvable theory
for which the S-matrices 
have been formulated. However, when a boundary is
present then the boundary 
S-matrices of the theory i.e.
the reflection factors, have 
not been completely  found. The bootstrap technique
does  not  uniquely  determine  the reflection factors.
Fortunately
 perturbation theory provides  the link between the
expressions for the reflection factors which come from the 
bootstrap equations and the boundary parameters.
Nevertheless, 
this method normally involves  complicated calculations.

In this paper the quantum reflection factor for the
$a_{1}^{(1)}$ affine
Toda field theory or sinh-Gordon model with integrable boundary
conditions has been studied in low order 
perturbation theory when $\sigma_{0} \neq \sigma_{1}$. It is
found that at
one loop order the quantum corrections to the classical reflection
factor of the model can be expressed in terms of  hypergeometric
functions for most of the related Feynman diagrams. Although there
is still some work  to do to calculate  the contributions of the
remaining diagrams, it is understood 
that the provided procedure and some formalisms may
be followed for them.    

 The 
calculations corresponding to the type II Feynman
diagram which are not carried out  in 
this paper, are more difficult than the others. 
In this case the 
two middle propagators are exactly the same and
this fact influences the 
difficulty  of the computations. However some formulae 
that have been presented here, 
could be helpful for the remaining diagram.
For example, consider  the contribution of the type II
(boundary-bulk) diagram:
  
\begin{eqnarray}
& &-2\beta^{2}(\sigma_{1}\coth x_{0}-\sigma_{0}\tanh x_{0})\int
\int \int
dt dt' dx G(x_{1},t_{1};x,t)G(x,t;0,t')
\nonumber\\
&
&\hspace{2.2in} \times
G(x,t;0,t')G(0,t';x_{2},t_{2})\sinh(\sqrt{2}\beta
\phi_{0}).
\end{eqnarray}
Now as far as the integration over $x$ is concerned we should obtain
the following integrals
\begin{equation}
\int_{-\infty}^{0}dx\,e^{i(k+k'-k_{1})x}
\sinh(\sqrt{2}\beta\phi_{0})
\coth^{n}2(x-x_{0}),
\end{equation}
where n=0,1,2,3. It is better to solve:
\begin{equation}          
\int_{-\infty}^{0}dx\,\exp{\left\{\tau +i(k+k'-k_{1})x\right\}}
\sinh(\sqrt{2}\beta\phi_{0})
\coth^{n}2(x-x_{0})
\end{equation}
in which $\tau$ is a small positive quantity and   will be
taken to zero later. In fact, the relation (8.3)  is very
similar
to the formula (A.1)  in Appendix A. So, following  the same
procedure that have been followed in Appendix A, one can find
the solution of  (8.2) when $n=0$ as:
  
\begin{eqnarray}
\int_{-\infty}^{0}&dx&e^{
i(k+k'-k_{1})x}\sinh(\sqrt{2}\beta
\phi_{0})
\nonumber\\
&=&\frac{1}{\sinh 2x_{0}}
\nonumber\\
& &-\frac{2(k+k'-k_{1})}{k+k'-k_{1}-2i}
\nonumber\\ 
& &\,\,\,\,\,\times \,\, e^{-2x_{0}}F(1,
\frac{i}{4}(k+k'-k_{1})+\frac{1}{2},\frac{i}{4}(k+k'-k_{1})+
\frac{3}{2},
e^{-4x_{0}}).
\end{eqnarray}
Then, the solutions  of  (8.2) for $n=1,2,3$ can be derived
 exactly in according to the Appendix A terms. But, this is not
 all of the problem. As we mentioned before, in type II
diagram double Green functions cause  the middle momenta to be
linked to each other in a complicated way and the calculations
become more intricate. 
Actually this diagram must
be studied in three cases depending on the interaction
vertices being  located in the bulk 
region or at the boundary. Moreover because of
the symmetry, the contribution of the type II 
(boundary-bulk) diagram is the same as the type II
(bulk-boundary) one.

When the boundary 
parameters are equal then, only the type I diagram is involved
in
the theory. As we mentioned before, 
in this special case \cite{C} the quantum
corrections
to
the classical 
reflection of the model have  been found and 
 Ghoshal's formula for the lightest breather is checked
perturbatively to $O(\beta^{2})$. 
In our case, we realised that the contribution of the
type I (bulk) reduces to the special case. 
Taking (7.7) and (7.19) expressions, if we put $\sigma_{0} =
\sigma_{1}$ then, we obtain the same result as  reference
\cite{C} and this is a   check on  our calculations.  
Moreover, when $\sigma_{0} \neq \sigma_{1}$ 
 the following expressions for $E$ and
$F$  in  Ghoshal's formula (2.8) have been conjectured
\cite{CC} to be:
\begin{equation}\label{conjecture}
E=(a_0+a_1)(1-B/2) \qquad F=(a_0-a_1)(1-B/2).
\end{equation}
So, it will be interesting  to check the above conjecture
after finding the contributions of the remaining 
diagrams and adding the 
results all together. This will lead to a deeper
understanding of the quantum integrability of the theory. However,
it is necessary to find   
  simplifications of the contributions when they add
among themselves in order to get  Ghoshal's formula. 

\resection {Acknowledgement}
We would like to thank E. Corrigan and P. Bowcock for encouragement,
discussions and suggestions, and the Ministry of Culture and Higher
Education of Iran for financial support.

\appendix 
\appsection{}

In this Appendix we obtain such integrals
\begin{equation}\label{A1}
S_{n}=\int_{-\infty}^{0}dx\,e^{(2+ik)x}
\sinh(\sqrt{2}\beta\phi_{0})
\coth^{n}2(x-x_{0})
\end{equation}
in which $n=0,1,2,3$,  $\phi_{0}$ is the background solution to
the
equation of field
so that $\sinh(\sqrt{2}\beta\phi_{0})$ is proportional to the 
bulk three point coupling which is given by 
\begin{equation}\label{A2}
\sinh(\sqrt{2}\beta\phi_{0})=2\cosh2(x-x_{0})\left(
\coth^{2}2(x-x_{0})-1\right).
\end{equation}
Let us  start with the simplest case when $n=0$.
Using  (\ref{A2}), we have 
\begin{equation}
S_{0}=-\int_{-\infty}^{0}e^{(2+ik)x}d\left(
\frac{1}{\sinh2(x-x_{0})}
\right)
\end{equation}
or,  after integration by parts 
\begin{equation}\label{A5}
S_{0}=\frac{1}{\sinh2x_{0}}
+2(2+ik)\int_{-\infty}^{0}dx\,e^{(2+ik)x}
\frac{1}{e^{2(x-x_{0})}-e^{-2(x-x_{0})}}.
\end{equation}
Now, according to  (\ref{boundary parameter}), if
$\sigma_{0}>\sigma_{1}$
then  $x_{0} \geq 0$. Otherwise, it is necessary to adjust the
background solution (\ref{background}) by shifting $x_{0}$ through
$i\pi/2$ in
order to be guaranteed that $x_{0}\geq 0$. The singularity in the
equation (\ref{background}) is unimportant provided $x_{0}$ is
positive. So,
from  now on it is assumed $\sigma_{0} \geq \sigma_{1}$. Therefore,  
$x_{0}$
is greater or equal to zero.  But, $x$ is less than zero,
so
$0<e^{4(x-x_{0})}<1$ and hence
\begin{equation}\label{A6}
\frac{1}{e^{2(x-x_{0})}-e^{-2(x-x_{0})}}=
-e^{2(x-x_{0})}\sum_{n=0}^{\infty}
e^{4n(x-x_{0})}.
\end{equation}
Substituting (\ref{A6}) in (\ref{A5}), we obtain
\begin{equation}
S_{0}=\frac{1}{\sinh2x_{0}}-2(2+ik)
\,e^{-2x_{0}}\int_{-\infty}^{0}dx
\,e^{(4+ik)x}\sum_{n=0}^{\infty}e^{4n(x-x_{0})}.
\end{equation}
Clearly, the  series is uniformly convergent so
the above relation becomes 
\begin{equation}
S_{0}=\frac{1}{\sinh2x_{0}}-2(2+ik)\,e^{-2x_{0}}
\sum_{n=0}^{\infty}e^{-4nx_{0}}
\int_{-\infty}^{0}dx\,e^{(4+4n+ik)x}.
\end{equation}
After integration over $x$, we obtain
\begin{equation}
S_{0}=\frac{1}{\sinh2x_{0}}+2i(2+ik)\,e^{-2x_{0}}
\sum_{n=0}^{\infty}\frac{e^{-4nx_{0}}}{k-(4+4n)i}.
\end{equation}
On the other hand, the above infinite series is  a
hypergeometric
function. That is 
\begin{equation}\label{hypergeometric 1}
\sum_{n=0}^{\infty}\frac{e^{-4nx_{0}}}{k-i(4+4n)}=
\frac{F(1,\frac{i}{4}k+1,
\frac{i}{4}k+2,e^{-4x_{0}})}
{k-4i}.
\end{equation}
Therefore, we get the following relation 
\begin{eqnarray}\label{A11}
S_{0}&=&\frac{1}{\sinh 2x_{0}}
\nonumber\\
& &-2\frac{k-2i}{k-4i}
 \,e^{-2x_{0}}\,F(1,
\frac{i}{4}k+1,\frac{i}{4}k+2,
e^{-4x_{0}}).
\end{eqnarray}
The hypergeometric function is defined by \cite{Arfken}
\begin{equation}
F(a,b,c,z)=\sum_{n=0}^{\infty}\frac{(a)_{n}
(b)_{n}}{(c)_{n}}z^{n}\hspace{.2in} c\neq 0, -1, -2, ...
\end{equation}  
where
\begin{equation}
(a)_{n}=\frac{\Gamma(a+n)}{\Gamma(a)}=a(a+1)...(a+n-1) 
\hspace{.2in}n=1,
2, 3, ...\,\,\,\,\,\,.
\end{equation}
The above series defines a 
function which is analytic when $|z|<1$. Also,  the derivatives  of the
hypergeometric function are  given by 
\begin{equation}
\frac{d^{n}}{dz^{n}}F(a,b,c,z)=\frac{(a)_{n}(b)_{n}}{(c)_{n}}
F(a+n,b+n,c+n,z).
\end{equation}

Next, for $n=1$ 
  using  (\ref{A2})  for the bulk three point coupling we have  
\begin{equation}\label{A18}
S_{1}=2\int_{-\infty}^{0}dx\,e^{(2+ik)x}\cosh 2(x-x_{0})
\coth2(x-x_{0})\left(\coth^{2} 2(x-x_{0})-1\right).
\end{equation}
On the other hand, if we differentiate    the  left hand side
of (\ref{A11})
with respect to $x_{0}$, which is given by
\begin{eqnarray}
& &\frac{\partial S_{0}}{\partial x_{0}}
 =-\int_{-\infty}^{0}dx\,e^{(2+ik)x}
\biggl(4\sinh 2(x-x_{0})\left(\coth^{2} 2(x-x_{0})-1\right)
\biggr.\nonumber\\
& & \biggl. 
\hspace{.9in}+8\cosh
2(x-x_{0})\coth2(x-x_{0})\left(1-\coth^{2}2(x-x_{0})\right)\biggr)  
\end{eqnarray}
and by comparing the above 
formula with (\ref{A18}), then the following
equation
may be derived
\begin{equation}\label{A20}
S_{1}=\frac{1}{4} \frac{\partial S_{0}}{\partial x_{0}}+
\int_{-\infty}^{0}dx\,e^{(2+ik)x}\,
\frac{1}{\sinh 2(x-x_{0})}.
\end{equation}
The second term in the above relation can be manipulated as before.
Moreover,  it is evident that 
we need to differentiate     the right
hand side of (\ref{A11}) which is equal to 
\begin{eqnarray}
 \frac{\partial S_{0}}{\partial x_{0}}
&=&-\frac{2\cosh 2x_{0}}{\sinh^{2} 2x_{0}}
\nonumber\\
& &+4\frac{k-2i}{k-4i}\,e^{-2x_{0}}\,
F(1,\frac{i}{4}k+1,
\frac{i}{4}k+2,e^{-4x_{0}})
\nonumber\\
&
&+8\frac{k-2i}{k-8i}\,e^{-6x_{0}}\,
F(2,\frac{i}{4}k+2,
\frac{i}{4}k+3,e^{-4x_{0}}).
\end{eqnarray} 
Finally, by substituting the 
  relation (A.17) in (\ref{A20}),  doing
the
computation of second term in the right-hand side of (\ref{A20}) and
after simplifying we
obtain 
\begin{eqnarray}
S_{1}&=&-\frac{\cosh2x_{0}}
{2\sinh^{2} 2x_{0}}\nonumber\\
& &+\frac{k}{k-4i}\,e^{-2x_{0}}\,
F(1,\frac{i}{4}k+1,
\frac{i}{4}k+2,e^{-4x_{0}})
\nonumber\\
& &+2\frac{k-2i}{k-8i}\,e^{-6x_{0}}\,
F(2,\frac{i}{4}k+2,
\frac{i}{4}k+3,e^{-4x_{0}}).
\end{eqnarray}

In the same way, we may derive   
(\ref{A1}) when n is equal to 2 or 3,
however
gradually the calculations become lengthy and we only write down the
results, that is
\begin{eqnarray}
S_{2}&=&\frac{2}{3}\,\frac{1}{\sinh
2x_{0}}+\frac{1}{6}\,
\frac{\cosh^{2}2x_{0}+1}{\sinh^{3}2x_{0}}
\nonumber\\
&
&-\frac{1}{3}\,\frac{5k-8i}{k-4i}\,e^{-2x_{0}}\,
F(1,\frac{i}{4}k+1,
\frac{i}{4}k+2,e^{-4x_{0}})
\nonumber\\
&
&-\frac{4}{3}\,\frac{2k-3i}{k-8i}\,e^{-6x_{0}}\,
F(2,\frac{i}{4}k+2,
\frac{i}{4}(k+3,e^{-4x_{0}})
\nonumber\\
&
&-\frac{8}{3}\,\frac{k-2i}{k-12i}\,e^{-10x_{0}}\,
F(3,\frac{i}{4}k+3,
\frac{i}{4}k+4,e^{-4x_{0}})
\end{eqnarray}
and
\begin{eqnarray}
S_{3} &=&-\frac{13}{24}\,\frac{\cosh 2x_{0}}
{\sinh^{2} 2x_{0}}-\frac{1}{24}\,\frac{\cosh^{3}2x_{0}+5\cosh
2x_{0}}{\sinh^{4} 2x_{0}}\nonumber\\
&
&+\frac{1}{6}\,\frac{7k-4i}{k-4i}\,e^{-2x_{0}}\,
F(1,\frac{i}{4}k+1,
\frac{i}{4}k+2,e^{-4x_{0}})
\nonumber\\
&
&+\frac{1}{3}\,\frac{13k-22i}{k-8i}\,e^{-6x_{0}}\,
F(2,\frac{i}{4}k+2,
\frac{i}{4}k+3,e^{-4x_{0}})
\nonumber\\
&
&+\frac{1}{3}\,\frac{18k-32i}{k-12i}\,e^{-10x_{0}}\,
F(3,\frac{i}{4}k+3,
\frac{i}{4}k+4,e^{-4x_{0}})
\nonumber\\
& &+4\frac{k-2i}{k-16i}\,e^{-14x_{0}}\,
F(4,\frac{i}{4}k+4,
\frac{i}{4}k+5,e^{-4x_{0}}).
\end{eqnarray}

\appsection{}

In this Appendix we find  the following integrals 

\begin{equation}
C_{n}=\int_{-\infty}^{0} dx \,e^{ikx} \cosh(\sqrt{2}\beta\phi_{0}) 
\coth^{n}2(x-x_{0}).
\end{equation}
Here, $\cosh(\sqrt{2}\beta\phi_{0})$ is proportional to the bulk four
point coupling and is given by

\begin{equation}
\cosh(\sqrt{2}\beta\phi_{0})=\left( 2\coth^{2}2(x-x_{0})-1\right).
\end{equation}
So, let us  calculate such integrals
\begin{equation}\label{B1}
I_{n}=\int_{-\infty}^{0} dx \,e^{ikx} \coth^{n}2(x-x_{0}),
\end{equation} 
where $n=1, 2, ..., 6$
. It is better to find the solution of the above integrals
when $n=1$. Considering the following inequality (see Appendix A) 

\hspace{2in} $$0<e^{4(x-x_{0})}<1$$ 
and therefore, in what follows we will use the expanded form of 
$\coth
2(x-x_{0})$ as
\begin{equation}\label{B2}
\coth 2(x-x_{0})=1-2\sum_{n=0}^{\infty}e^{4n(x-x_{0})}.
\end{equation} 
It turns out to be simple if we consider this integral
\begin{equation}\label{B3}
\int_{-\infty}^{0} dx \,e^{(\tau+ik)x} \coth 2(x-x_{0}),
\end{equation}
where $\tau$ is a positive constant 
quantity which will be taken to zero at the end
of the calculation. 
Moreover, by using  (\ref{B2}), then (\ref{B3})
becomes 
\begin{equation}
\int_{-\infty}^{0} dx \,e^{(\tau+ik)x} 
-2\int_{-\infty}^{0} dx \sum_{n=0}^{\infty}e^{4n(x-x_{0})} 
\,e^{(\tau+ik)x} 
\end{equation}
and regarding the arguments in Appendix A, we may evaluate the above
relation to obtain
\begin{equation}
-\frac{i}{k-i\tau}+2\sum_{n=0}^{\infty}
\frac{i}{k-(\tau+4n)i}\,\,e^{-4nx_{0}}.
\end{equation}
Now, we can  write down the desired  result, that
is, 
\begin{equation}
I_{1}=-\frac{i}{k}+2i\sum_{n=0}^{\infty}\frac{e^{-4nx_{0}}}{k-4ni}. 
\end{equation}
On the other hand, the above series is equal to a hypergeometric
function 
\begin{equation}\label{hypergeometric 2}
\sum_{n=0}^{\infty}\frac{e^{-4nx_{0}}}{k-4ni} 
=\frac{1}{k} F(1,\frac{i}{4}k,\frac{i}{4}k+1,e^{-4x_{0}})
\end{equation}
and finally we find this formula
\begin{equation}\label{B11}
I_{1}=-\frac{i}{k}+\frac{2i}{k}
F(1,\frac{i}{4}k,\frac{i}{4}k+1,e^{-4x_{0}}).
\end{equation}

Now, let us  compute  (\ref{B1})  when $n=2$
and in order to solve it,  
it is sufficient to differentiate 
 both sides  of  
(\ref{B11}) with respect to $x_{0}$ to obtain
\begin{equation}
I_{2}=-\frac{i}{k}-\frac{4i}{k-4i}\,e^{-4x_{0}}\,
F(2,\frac{i}{4}k+1,\frac{i}{4}k+2,e^{-4x_{0}}).
\end{equation}  

We can follow a similar method to get higher order forms of 
(\ref{B1})
 which we need   in this paper so, it is appropriate to write
down all of
them i.e.
\begin{eqnarray}
I_{3}&=&-\frac{i}{k}+\frac{2i}{k}
F(1,\frac{i}{4}k,\frac{i}{4}k+1,e^{-4x_{0}})
\nonumber\\
&
&\frac{4i}{k-4i}\,e^{-4x_{0}}\,F(2,\frac{i}{4}k+1,
\frac{i}{4}k+2,e^{-4x_{0}})
\nonumber\\
&
&\frac{8i}{k-8i}\,e^{-8x_{0}}\,
F(3,\frac{i}{4}k+2,\frac{i}{4}k+3,e^{-4x_{0}}),
\end{eqnarray}  
\begin{eqnarray}
I_{4}&=&-\frac{i}{k}-\frac{8i}{k-4i}
\,e^{-4x_{0}}\,F(2,\frac{i}{4}k+1,\frac{i}{4}k+2,e^{-4x_{0}})
\nonumber\\
&
&-\frac{16i}{k-8i}\,e^{-8x_{0}}\,
F(3,\frac{i}{4}k+2,\frac{i}{4}k+3,e^{-4x_{0}})
\nonumber\\
&
&-\frac{16i}{k-12i}\,e^{-12x_{0}}\,
F(4,\frac{i}{4}k+3,\frac{i}{4}k+4,e^{-4x_{0}}),
\end{eqnarray}
\begin{eqnarray}
I_{5}&=&-\frac{i}{k}+\frac{2i}{k}
F(1,\frac{i}{4}k,\frac{i}{4}k+1,e^{-4x_{0}})
\nonumber\\
&
&\frac{8i}{k-4i}\,e^{-4x_{0}}\,
F(2,\frac{i}{4}k+1,\frac{i}{4}k+2,e^{-4x_{0}})
\nonumber\\
&
&\frac{32i}{k-8i}\,e^{-8x_{0}}\,
F(3,\frac{i}{4}k+2,\frac{i}{4}k+3,e^{-4x_{0}})
\nonumber\\
&
&\frac{48i}{k-12i}\,e^{-12x_{0}}\,
F(4,\frac{i}{4}k+3,\frac{i}{4}k+4,e^{-4x_{0}})
\nonumber\\
&
&\frac{32i}{k-16i}\,e^{-16x_{0}}\,
F(5,\frac{i}{4}k+4,\frac{i}{4}k+5,e^{-4x_{0}})
\end{eqnarray}
and
\begin{eqnarray}
I_{6}&=&-\frac{i}{k}
-\frac{12i}{k-4i}\,e^{-4x_{0}}\,
F(2,\frac{i}{4}k+1,\frac{i}{4}k+2,e^{-4x_{0}})
\nonumber\\
&
&-\frac{48i}{k-8i}\,e^{-8x_{0}}\,
F(3,\frac{i}{4}k+2,\frac{i}{4}k+3,e^{-4x_{0}})
\nonumber\\
&
&-\frac{112i}{k-12i}\,e^{-12x_{0}}\,
F(4,\frac{i}{4}k+3,\frac{i}{4}k+4,e^{-4x_{0}})
\nonumber\\
&
&-\frac{128i}{k-16i}\,e^{-16x_{0}}\,
F(5,\frac{i}{4}k+4,\frac{i}{4}k+5,e^{-4x_{0}})
\nonumber\\
&
&-\frac{64i}{k-20i}\,e^{-20x_{0}}\,
F(6,\frac{i}{4}k+5,\frac{i}{4}k+6,e^{-4x_{0}}).
\end{eqnarray}

\end{document}